\documentclass{article}

\usepackage{jcappub}

\usepackage{mathtools}
\usepackage{pifont}
\usepackage{graphicx}
\usepackage[compat=1.1.0]{tikz-feynman}
\usepackage{feynmp}

\newcommand{\gs}{g_\star}
\newcommand{\gss}{g_{\star s}}

\newcommand{\rGW}{\rho_\text{GW}}
\newcommand{\rR}{\rho_R}

\newcommand{\mueff}{\mu_\text{eff}}
\newcommand{\yeff}{y_\text{eff}}
\newcommand{\Eom}{E_\omega}

\title{Spectrum of high-frequency gravitational waves from graviton bremsstrahlung by 
the decay of inflaton: case with polynomial potential}
\author[a]{Yiheng Jiang}
\author[b]{and Teruaki Suyama}

\affiliation[a,b]{Department of Physics, Institute of Science Tokyo, 2-12-1 Ookayama, Meguro-ku,
Tokyo 152-8551, Japan}

\abstract{We study the generation of high-frequency gravitational waves (GWs) through graviton bremsstrahlung during the decay of inflaton in the post-inflationary universe, focusing on scenarios with a polynomial inflaton potential.
Two main reheating channels are considered: decays into bosons (spin 0) and fermions (spin 
$\frac{1}{2}$). We compute the resulting GW spectra from three-body decays, where the inflaton decays into a pair of daughter particles and a graviton. 
We numerically compute the GW spectra for various polynomial exponents by employing two distinct approaches: one treating the inflaton as a collection of rest particles and the other treating it as a coherently oscillating classical field.
In the former approach, only gravitons with energies below half the inflaton mass are produced, while the latter allows for the production of gravitons with arbitrarily high energies when the potential exponent is 4 or greater.
This difference arises because the inflaton's oscillations are no longer 
described by a single harmonic mode but instead consist of infinitely many harmonic modes 
with different frequencies.
As a result, the GW spectrum exhibits multiple peaks, 
with these peaks being less pronounced for higher powers of the potential. 
We also examine the dependence of the GW spectrum on the coupling constant between the inflaton and daughter particles. 
Our findings suggest that future GW detectors targeting GWs in the {\rm GHz} band, such as resonant cavities, may have the capability to detect these signals, offering potential insights into the reheating phase of the early universe.}

\begin{document}
        \begin{flushright}
	\end{flushright}
	\maketitle


\section{Introduction}
Inflation, the phase of accelerated expansion in the early universe, not only solves the puzzles of Big Bang cosmology
such as the horizon and flatness problems, but also provides a natural explanation for the origin
of primordial perturbations ~(e.g., ~\cite{Baumann:2022mni}).
After the end of inflation when the kinetic energy becomes equal to the potential energy, 
inflaton oscillates coherently around the bottom of the potential. 
During this oscillation phase, inflaton gradually decays and produces other relativistic particles.
At around the lifetime of inflaton, radiation consisting of decay products 
dominates over the inflaton and this time, which marks the beginning of the radiation
dominated era, is known as reheating.

Contrary to the deep inflation era where produced primordial perturbations are
directly compared with observations, the physical phenomena occurring during the oscillating phase 
of inflaton are challenging to probe experimentally. 
Among the possible observables, gravitational waves (GWs) hold great potential for providing information about reheating, as GWs generated in the early Universe can contribute to the present-day stochastic gravitational wave background (SGWB), which is an important target for future GW detectors.
During the oscillation stage of inflaton, the following mechanisms capable of generating 
high-frequency GW signals have been recently studied.
One is GWs produced from the three-body decay of the inflaton in which, if we view the oscillating 
inflaton as a group of particles at rest, an inflaton decays into a pair of daughter particles (either
bosons or fermions) and a graviton~\cite{Nakayama_2019,Barman_2023,Barman:2023rpg,kanemura2024gravitationalwavesparticledecays,
Bernal:2023wus,ghoshal2023bremsstrahlunghighfrequencygravitationalwave,tokareva2024gravitationalwavesinflatondecay,choi2024gravitationalwavesourceddecay,Huang_2019,hu2024gravitationalwaveprobeplanckscale,xu2024ultrahighfrequencygravitationalwaves}. 
This three-body decay, also known as graviton bremsstrahlung, 
inevitably occurs as long as inflaton interacts with other fields since such interaction term 
is always coupled to gravity.
GWs can also arise directly from the inflaton without producing other daughter particles, 
which may be interpreted as inflatons annihilating into a pair of gravitons solely through gravitational coupling~\cite{Ema_2015,Ema_2016,ema2020highfrequencygravitoninflatonoscillation,choi2024minimalproductionpromptgravitational}. Additionally, GWs can be generated from the thermal fluctuations of the standard model (SM) plasma~\cite{Ghiglieri_2015,Ghiglieri_2020,Ghiglieri_2024,drewes2023upperboundthermalgravitational} or from the parametric oscillation process during preheating~\cite{Khlebnikov_1997,Easther_2006,Dufaux_2007,Bethke_2013,Figueroa_2017}. 

In this article, we focus on the gravitational waves generated by the graviton bremsstrahlung. 
The first study on graviton bremsstrahlung during the oscillation period of inflaton was conducted in \cite{Nakayama_2019},
in which the authors calculated the GW spectrum both in the case of bosonic decay and fermionic decay, assuming that the
inflaton potential around the bottom takes a quadratic form. 
In \cite{Barman_2023,Barman:2023rpg}, subsequent works investigated 
graviton bremsstrahlung for decay products with different spins and different shapes of the inflaton potential. 
In \cite{kanemura2024gravitationalwavesparticledecays}, the authors examined the process of secondary decay, 
where the initial decay of inflaton produces heavy particles and the three-body decay of the heavy particles
significantly increases the GW spectrum. 
In \cite{Bernal:2023wus}, gravitons produced during the scattering of inflatons were studied. Additionally, some studies have investigated GWs produced through bremsstrahlung during lepton generation \cite{ghoshal2023bremsstrahlunghighfrequencygravitationalwave}, 
while others have examined the modifications to GW production from bremsstrahlung within the framework of effective gravitational theories \cite{tokareva2024gravitationalwavesinflatondecay}. 
Other scenarios based on the graviton bremsstrahlung include the decays of particles resulting from primordial black hole evaporation \cite{choi2024gravitationalwavesourceddecay} and the decays of super-heavy
particles in the kination phase after inflation \cite{Inui:2024wgj}.

Although the simplest form of the inflaton potential around the bottom is quadratic,
this assumption could be incorrect in reality.
For instance, in the $\alpha$-attractor T model \cite{Kallosh_2013}, the potential around the bottom becomes $V(\phi) \sim \phi^k$
with $k \ge 2$.
While the universe expands like matter-dominated universe during the oscillation phase if the exponent $k$ is $2$,
this is not the case if $k$ is different from $2$ because the equation of state parameter depends on $k$.
In \cite{Barman:2023rpg, Bernal:2023wus}, it was demonstrated that the information of the exponent $k$ 
is encoded in the GWs produced by the graviton bremsstrahlung and future GW detectors may be able to 
probe the shape of the inflaton potential in the oscillation period.
Thus, there is a strong motivation to study the GW production by the graviton bremsstrahlung in detail and
establish theoretical understanding and framework about this physical process.
 
In the literature, studies on graviton bremsstrahlung typically assume that the inflaton consists of a large number of particles at rest, with its mass given by the second derivative of the potential. 
The decay of the inflaton is then calculated using the standard Quantum Field Theory (QFT) in Minkowski space \footnote{The period of the oscillation of the inflaton is much shorter than the Hubble time. Thus, it is a good approximation to
ignore the effect of the cosmic expansion and to compute the transition amplitude assuming the background spacetime
is Minkowski.}. 
However, given that the inflaton field is a homogeneous classical field,
in principle, the decay rate of the inflaton and the GW spectrum should be computed by treating the inflaton
as the classical oscillating field.
In this paper, we call the calculations based on the former picture {\it particle picture}
and the latter picture {\it classical field picture}, respectively.
Although the computations based on the classical field picture are done for the two-body decay \cite{Ichikawa:2008ne, Garcia:2020wiy, Barman:2023rpg, Bernal:2023wus},
calculations for the graviton bremsstrahlung are done only in the particle picture 
and the ones
based on the classical field picture are not found in the literature.
Technically, the time variation of the inflaton can be split into slowly-decaying part $\phi_0 (t)$ due to the cosmic
expansion and the rapidly oscillating part $\mathcal{P}(t)$ as $\phi(t) = \phi_0(t)\mathcal{P}(t)$~\cite{Garcia:2020wiy}. 
If the exponent $k$ is different from $2$, $\mathcal{P} (t)$ consists of the superposition 
of the fundamental oscillation mode and higher frequency modes.
While the particle picture accounts only for the fundamental mode, the classical field picture includes the contribution
from all the modes, which will result in difference of the GW spectrum.

In this paper, we compute the GW spectrum based on both the two pictures and compare their differences. 
The characteristic signals in the gravitational wave spectrum caused by the superposition of the inflaton’s 
oscillation modes are also studied. 
Another new point which is missing in the literature in the context of the computations of the GW spectrum
from the graviton bremsstrahlung is that we solve the time evolution of both inflaton and the GW spectrum
fully numerically without making approximations such as sudden decay approximation for the inflaton. 

This paper is organized as follows.
In section \ref{sec:headings}, we briefly review the post-inflationary dynamics and the approximate analytical solution for the radiation density. 
In section \ref{Sec:Bremsstrahlung}, we calculate the three-body differential decay rates in the two pictures. 
In section \ref{sec:numerical-results}, the numerical results are given. 
Finally, we summarize our results and draw conclusions in Section \ref{sec:conclusion}.

\section{Post-inflationary dynamics}
\label{sec:headings}
\numberwithin{equation}{section}

\subsection{The model}
After inflation ends, the Universe enters the stage of rapid oscillations of the inflaton. 
Because of the coupling between the inflaton and the matter field, the energy of the inflaton is transferred to the SM bath as the inflaton decays. 
In this paper, we consider two decay channels; decay into a pair of bosons (denoted by $b$)/fermions (denoted by $f$).
We assume that masses of daughter particles are negligibly small and we treat
those particles as massless.
The action that describes dynamics of inflaton $\phi$ and the decay processes is given by
\begin{equation}
\mathcal{S}=\int d^4 x\sqrt{-g}\left[\frac{1}{16\pi G_N} R+\frac12 (\partial_\mu \phi)^2-V(\phi)+\frac12 (\partial_\mu b)^2-{\bar \psi} \gamma^\mu \nabla_\mu \psi +\mathcal{L}_\text{int}\right].
\end{equation}
Here $V(\phi)$ is the potential of the inflaton and $\mathcal{L}_\text{int}$ which 
represents the interaction between inflaton and the matter field is given by
\begin{equation}\label{eq:int4}
\mathcal{L}_\text{int}= \mu\phi b b +y\phi \bar{f} f.
\end{equation}
$\mu$ and $y$ are coupling constants.
Without a loss of generality, we can assume that the potential is minimum at $\phi=0$. 
In the oscillation phase of inflaton, only a small range of $\phi$ around $\phi=0$ is relevant. 
As it is explained in the Introduction, we assume that 
the potential in this regime takes a polynomial form
\begin{equation}\label{eq:int3}
V(\phi)=\lambda M_p^4 \left(\frac{\phi}{M_p}\right)^k, \quad\phi\ll M_p
\end{equation}
Here $M_p= 1/\sqrt{8\pi G_N}\simeq2.4\times 10^{18} $ GeV is the reduced Planck mass, and $\lambda$ is a constant that parametrizes the energy scale of the potential after inflation.
Since the potential associated with the inflationary phase may differ from the post-inflationary potential, the value of 
$\lambda$ can only be determined from CMB observations, such as those by the PLANCK mission, once the inflationary potential is specified. However, because the specific form of the inflationary potential is not directly relevant to our analysis (i.e., the assumption in Eq.~(\ref{eq:int4}) suffices), we refrain from adopting a specific potential. Instead, we simply use $\lambda=10^{-11}$ as a representative value in our numerical calculations.
In Sec.~\ref{sec:numerical-results}, we provide an approximate relation that describes how the GW spectrum changes with variations in 
$\lambda$. 
This relation enables a straightforward determination of the GW spectrum for different values of $\lambda$.

The action above contains interactions between gravitational field and the other fields.
We define the perturbative gravitational field $h_{\mu \nu}$ by 
$g_{\mu\nu}\simeq \eta_{\mu\nu}+\frac2{M_p} h_{\mu\nu}$ and only consider
the interactions between $h_{\mu \nu}$ and the other fields at the lowest order 
in $h_{\mu \nu}$ because the higher order interactions give only negligible contributions.
Then the interaction terms including $h_{\mu \nu}$ are given by~\cite{Choi_1995}.
 \begin{equation}
 \sqrt{-g}\mathcal{L}_\text{int, h}= -\frac1{M_p}h_{\mu\nu}T^{\mu\nu}
 \end{equation}
where $T^{\mu\nu}$ denotes the energy-momentum tensor of the matter fields.

\subsection{Dynamics of inflaton without the decay}
Let us briefly review some properties of the oscillations of the inflaton in the
absence of decay, which is a good approximation at times much before the lifetime.
Adopting the inflaton potential given by Eqs.~\eqref{eq:int3}, 
equation of motion for the inflaton is given by
\begin{equation}\label{eq:int6}
\ddot{\phi}+3H\dot{\phi}+V'(\phi)=0,~~~~~~~~V'(\phi)=k\lambda M_p^{4-k} \phi^{k-1}. 
\end{equation}
When there is a large hierarchy between the Hubble time and the oscillation period,
it is legitimate to separate the time dependence of $\phi$ into a product of
amplitude part $\phi_0$ which slowly decays due to the cosmic expansion and the purely oscillatory part $\mathcal{P}$ as
\begin{equation}
\phi(t)=\phi_0(t)\mathcal{P}(t).
\end{equation}
By definition, the amplitude of $\mathcal{P}$ is normalized to unity.
Plugging this decomposition into the equation of motion and picking up the
leading order terms, we obtain two equations
for $\mathcal{P}$ and $\phi_0$ as
\begin{align}
&{\ddot {\mathcal{P}}}+k\lambda M_p^{4-k} \phi_0^{k-2} {\mathcal{P}}^{k-1}=0, 
\label{eq-P-no-friction}
\\
&{\ddot \phi_0}+3H {\dot \phi_0}+k\lambda M_p^{4-k} \phi_0^{k-1} \langle \mathcal{P}^{k-1} \rangle =0.
\label{eq-phi0-no-friction}
\end{align}
Here, $\langle...\rangle$ means averaging over one oscillation and it should be
understood that $\phi_0$ can be treated as constant during one period.
Since there is no friction term in Eq.~(\ref{eq-P-no-friction}), $\mathcal{P}$ is purely
a periodic function with period $T$ given by
\begin{equation}
T=\frac{4\sqrt{\pi}}{k\sqrt{2\lambda M_p^{4-k} \phi_0^{k-2}}} 
\frac{\Gamma (\frac{1}{k})}{\Gamma (\frac{1}{2}+\frac{1}{k})}, 
\end{equation}
where we have used a formula $\int_0^1 \frac{dx}{\sqrt{1-x^k}}=\frac{\sqrt{\pi} \Gamma (\frac{1}{k})}{k\Gamma (\frac{1}{2}+\frac{1}{k})}$.
Equivalently, in terms of the effective mass defined by the second derivative
of the potential evaluated at $\phi_0$ as
\begin{equation}\label{eq:int14}
m_\phi^2(t)\equiv V''(\phi_0),
\end{equation}
the angular frequency $\omega =\frac{2\pi}{T}$ is written as
\begin{equation}
\label{def-angular-frequency}
\omega=m_\phi\sqrt{\frac{\pi k}{2(k-1)}}\frac{\Gamma(\frac12+\frac1k)}{\Gamma(\frac1k)}.
\end{equation}
By the periodic nature of $\mathcal{P}$, 
$\mathcal{P}$ can be expressed as a superposition of sinusoidal functions with
multiple of the fundamental frequency $\omega$ as
\begin{equation}
\mathcal{P}(t)=\sum_{n=-\infty}^{\infty}\mathcal{P}_n e^{-in\omega t}.\label{eq:int2.12}
\end{equation}
Since $\mathcal{P}$ is a real function, 
$\mathcal{P}_{-n}=\mathcal{P}_n$ is imposed. 
From the condition $\langle \mathcal{P} \rangle =0$, we also have $\mathcal{P}_0=0$.
If $k=2$, the motion of $\mathcal{P}$ is that of the harmonic oscillator and only
$n=\pm 1$ modes exist.
If $k>2$, $\mathcal{P}_n$ is non-zero in general.

For later purpose, we introduce the energy density and pressure of the inflaton 
which can be derived from the energy-momentum tensor as
\begin{equation}
\rho_\phi=\frac12 \dot{\phi}^2+V(\phi),~~~~~~~~~
\quad P_\phi=\frac12 \dot{\phi}^2-V(\phi).
\end{equation}
The equation of motion of $\phi$ (\ref{eq:int6}) yields evolution equation
for $\rho_\phi$ as
\begin{equation}
\dot{\rho_\phi}+3H(\rho_\phi+P_\phi)=0.
\end{equation}
This is an exact equation and ${\dot \rho_\phi}$ oscillates on the time scale $T$.
Since we are not interested in such rapid time variation whose effect 
disappears after averaging over $T$, let us take the average of the above equation;
\begin{equation}
\frac{d}{dt} \langle \rho_\phi \rangle +3H \langle \rho_\phi+P_\phi \rangle=0.
\end{equation}
To evaluate the second term, we multiply Eqs.~\eqref{eq:int6} by $\phi$ 
and average over one oscillation. This gives the following relation
\begin{equation}
\langle \phi\ddot{\phi}+3H\phi\dot{\phi}+\phi V'(\phi)\rangle=-\langle\dot{\phi}^2\rangle+\langle \phi V'(\phi)\rangle=0,
\end{equation}\label{eq:int2.16}
from which we can relate $\langle P_\phi \rangle$ to $\langle \rho_\phi \rangle$ as
\begin{equation}
\langle P_\phi \rangle =\frac{k-2}{k+2} \langle \rho_\phi \rangle.
\end{equation}
Thus, the equation of state parameter is $w_\phi=\frac{k-2}{k+2}$.
With this parameter, time evolution of $\langle \rho_\phi \rangle$ becomes
\begin{equation}
\frac{d}{dt} \langle \rho_\phi \rangle +3H (1+w_\phi) \langle \rho_\phi \rangle=0.
\end{equation}
We can immediately integrate this equation and obtain
\begin{equation}\label{eq:int20}
\langle \rho_\phi \rangle =\rho_{\rm end}\left( \frac{a_\text{end}}{a} \right)^{\frac{6k}{k+2}},
\end{equation}
where the subscript "end" means that the corresponding quantity is evaluated
at the time of the end of inflation.
In what follows, we only use the averaged $\rho_\phi$ and
we omit the average symbol $\langle \rangle$ for notational simplicity.

\subsection{Inclusion of the decay}

Let us next include the decay of the inflaton and see how it changes the dynamics of
inflaton.
In the computations of the decay rate,
we ignore the effects due to the cosmic expansion and work in the QFT in Minkowski spacetime, which is a good approximation
when the oscillation period of inflaton is much shorter than the Hubble time.
As it is explained at the beginning of this section, we consider
spin $0$ bosons and $\frac{1}{2}$ fermions as decay products.
In addition to the spontaneous decay, inflaton decays via stimulated emission
if the daughter particles are bosons and the decay is blocked by the Pauli's
exclusion principle if the daughter particles are fermions.
In this paper, we assume the decay is in the perturbative regime and only 
consider the spontaneous emissions of daughter particles.
Then, the energy density of daughter particles transferred from inflaton during 
a short time interval $[t, t+\Delta t]$ is proportional to the energy density
of the inflaton and $\Delta t$; $\Delta \rho_R \propto \rho_\phi \Delta t$.
We define the decay rate of the inflaton $\Gamma_\phi$ as
\footnote{The additional factor $1+w_\phi$ in front of $\Gamma_\phi$ in Eq.~(\ref{def-Gammaphi}) has been introduced to
make the expression of the EoM of inflaton simple~\cite{PhysRevD.28.1243};
\begin{equation}
\ddot{\phi}+\left(3 H+\Gamma_\phi\right) \dot{\phi}+V^{\prime}(\phi)=0.
\end{equation}}
\begin{equation}
\label{def-Gammaphi}
\Delta \rho_R =(1+w_\phi) \Gamma_\phi \rho_\phi \Delta t.
\end{equation}
Explicit expression of $\Gamma_\phi$ will be given in the next section.
By the conservation of the total energy, which is the sum of the energy density
of inflaton and that of radiation originating from daughter particles, 
this is equal to the loss of the energy density of inflaton during the same time interval. 
Thus, the time evolution of the energy density of inflaton is given by
\begin{equation}
\dot{\rho}_\phi+3H(1+w_\phi)\rho_\phi =-\Gamma_\phi(1+w_\phi)\rho_\phi.
\end{equation}
Correspondingly, the evolution equation of radiation energy density 
becomes
\begin{equation}\label{eq:int18}
\dot{\rho_R}+4H\rho_R=\Gamma_\phi(1+w_\phi)\rho_\phi.
\end{equation}
The total energy satisfies the Friedmann equation,
\begin{equation}
H^2=\frac{1}{3M_p^2} (\rho_\phi+\rho_R).
\end{equation}

\section{Bremsstrahlung production of gravitational waves}
\label{Sec:Bremsstrahlung}
In this section, we provide the expressions of the decay rate of inflaton both for the
two-body decay and the three-body decay.

\subsection{Bosonic reheating}
\numberwithin{equation}{section}
For the bosonic decay, the two-body decay rate from the interaction term Eqs.~\eqref{eq:int4} based on the classical field picture was obtained in \cite{Ichikawa:2008ne, Garcia:2020wiy} as,  
\begin{equation}
\Gamma_\phi^{1\to2}(a)=\frac{\mu_{\text{eff}}^2}{8\pi m_\phi(a)}
\end{equation}\label{eq:int3.1}
Here $\mu_{\text{eff}}$ is the effective coupling constant defined in such a way that
it gives the same expression for the decay rate as the case of $k=2$;
\begin{equation}
\label{def-mueff}
\mu_{\text{eff}}^2(k)=(k+2)(k-1)\frac{\omega}{m_\phi}\left[\sum_{n=1}^{\infty}n|\mathcal{P}_n|^2 \right]\mu^2.
\end{equation}
By definition, $\mu_{\rm eff} =\mu$ for $k=2$.
For $k>2$, higher frequency modes contribute to the decay and the precise value
of $\mu_{\rm eff}$ can be obtained only numerically. 

At this stage, it is suggestive to rewrite the decay rate (\ref{eq:int3.1}) as
\begin{equation}
\label{eq:int3.4}
\Gamma_\phi^{1\to2}(a)=\sum_{n=1}^{\infty}b_n\Gamma_{\phi_n}^{1\to2}(a).
\end{equation}
Here $\Gamma_{\phi_n}^{1\to2}(a)$ defined by
\begin{equation}
\Gamma_{\phi_n}^{1\to2}(a)\equiv\frac{\mu^2}{8\pi n\omega(a)}
\end{equation}
represents the decay rate of an {\it inflaton} when it has a mass $n\omega$.
The coefficient $b_n$ is defined by
\begin{equation}
\label{def-bn}
b_n\equiv(k+2)(k-1)\left(\frac{\omega}{m_\phi}\right)^2n^2|\mathcal{P}_n|^2.
\end{equation}
Obviously, $b_n$ is positive. 
As it is proved in Appendix~\ref{Sec:bn}, $b_n$ satisfies the following key identity
\begin{equation}
\sum_{n=1}^{\infty}b_n=1.
\end{equation}
This relation allows us to identify $b_n$ as a fraction of massive particles having mass $n\omega$.
From these expressions, we can interpret that coherent oscillations of inflaton
is a collection of an infinite tower of massive particles where the particles 
labeled with $n$ have a mass $n\omega$, constitute a fraction $b_n$,
and decay into bosons $b$ with the decay rate $\Gamma_{\phi_n}^{1\to2}(a)$ computed 
based on the standard QFT.


\begin{figure}[t]
    \centering
    \includegraphics[width=.8\textwidth]{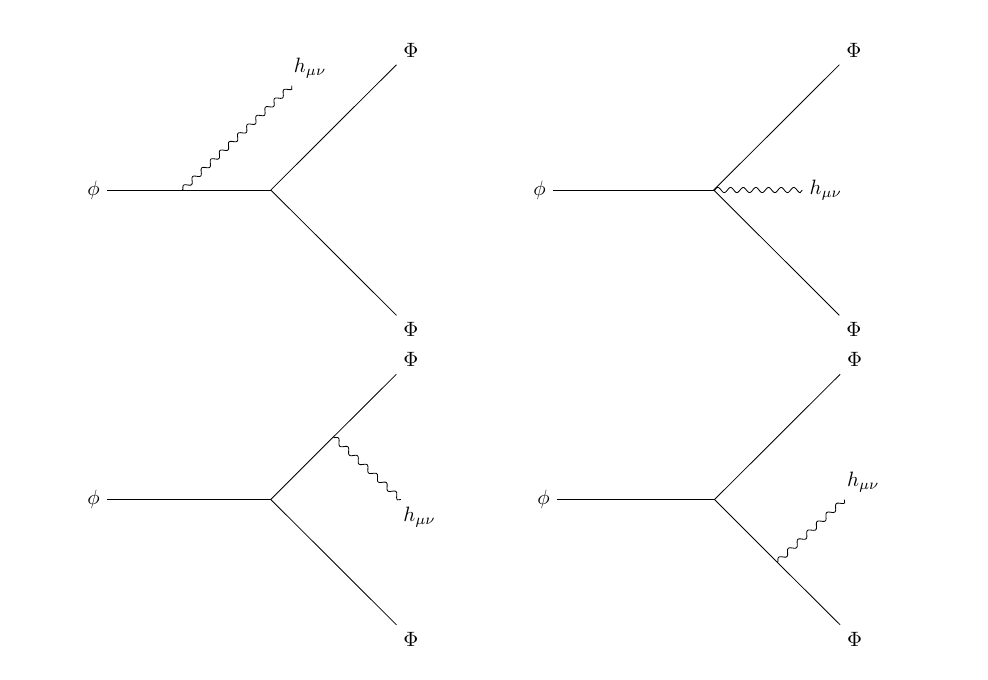}~~~
    \caption{Feynman diagrams for an inflaton decay into a pair of particles $\Phi$ and one graviton. The particles $\Phi$ can be bosons or fermions. This three-body decay process is called graviton bremsstrahlung.}
    \label{fig:Feynman diagrams}
\end{figure}

As for the graviton bremsstrahlung, we can understand the process in terms of the
Feynman diagrams in the particle picture, as shown in Fig.~\ref{fig:Feynman diagrams}. 
Since $h_{\mu \nu}$ is only coupled to terms containing spatial derivative of $\phi$
and the inflaton in the initial state have zero momentum,
transition amplitude of the top left diagram identically vanishes. 
The top right diagram also gives vanishing transition amplitude because of the
traceless nature of $h_{\mu \nu}$.
Thus, only the bottom two diagrams contribute to the decay process.
As it is explained in the Introduction, production rate of gravitons
due to the graviton bremsstrahlung has been computed based on the particle picture
using the Feynman diagrams in Fig.~\ref{fig:Feynman diagrams}
by assigning the effective coupling constant 
Eq.~(\ref{def-mueff}) defined for the two-body decay for the vertices in the diagrams
and then computing the transition amplitude in the standard matter \cite{Nakayama_2019, Barman:2023rpg}. 
The resultant differential decay rate is given by \cite{Nakayama_2019}
\begin{equation} \label{eq:dGdE_bos}
	\frac{d\Gamma_\phi^{1\to 3}}{dE_\omega} = \frac{1}{32\,\pi^3} \left(\frac{\mueff}{M_P}\right)^2 \frac{(1 - 2 x)^2}{4 x}\,, ~~~~~\quad 0<x<\frac12.
\end{equation}
Here $E_\omega$ is the energy of graviton and $ x=\frac{E_\omega}{m_\phi}\,$. 
Production of gravitons with energies larger than $\frac{m_\phi}{2}$ is forbidden kinematically.

Let us next address the differential decay rate in the classical field picture in which case the initial state is a classical field with no particles. 
In this picture, we substitute 
$\phi(t)=\phi_0(t)\sum_{n=-\infty}^{\infty}\mathcal{P}_n e^{-in\omega t}$ into the interaction term. 
By doing this, operators in that term are the scalar field $b$ only, but now the term has 
a time dependent coefficient. 
The transition amplitude can then be computed by using this interaction term
with no particles as the initial state and a pair of bosonic particles and a graviton
as the final state.
The intermediate calculations are given in Appendix~\ref{Sec:Decay Rate} and 
here we only give the final expression;
\begin{equation}\label{eq:int3.7}
	\frac{d\Gamma_\phi^{1\to 3}}{d\Eom} = \frac{1}{32\,\pi^3} \left(\frac{\mu}{M_P}\right)^2 (k+2)(k-1)\left(\frac{\omega}{m_{\phi}}\right)^2\left[\sum_{n=1}^{\infty} n^2 |\mathcal{P}_n|^2\left(\frac{(1 - 2 x_n)^2}{4 x_n} \right)\right],\quad 0<x_n<\frac12
\end{equation}
Here $x_n \equiv \frac{E_\omega}{n\omega}\,$. 
Clearly, while the production of gravitons with energy larger than $\frac{m_\phi}{2}$ is absent
in the decay rate (\ref{eq:dGdE_bos}) derived based on the particle picture, 
gravitons with arbitrarily large energy can be produced 
in the classical field picture.
This difference arises because the classical field picture accounts for all the
oscillation modes and the gravitons originating from the $n$-th mode have energy
in the kinematically allowed range $0< E_\omega < \frac{n\omega}{2}$.

 
 \begin{figure}[t]
		\centering
		\includegraphics[scale=0.65]{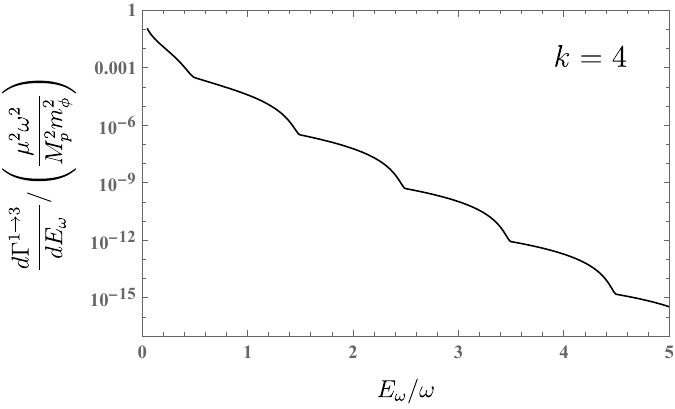}
		\includegraphics[scale=0.65]{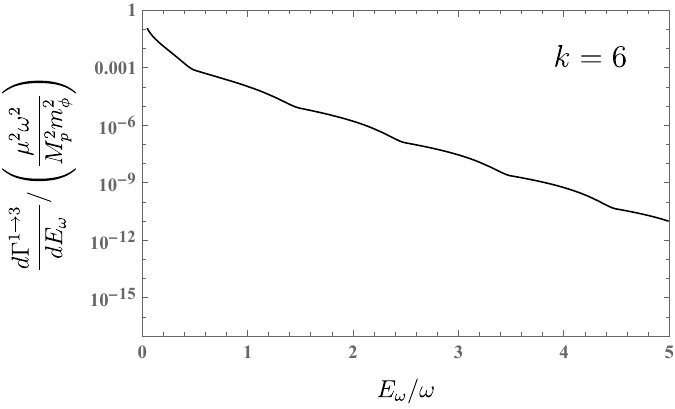}
            \includegraphics[scale=0.65]{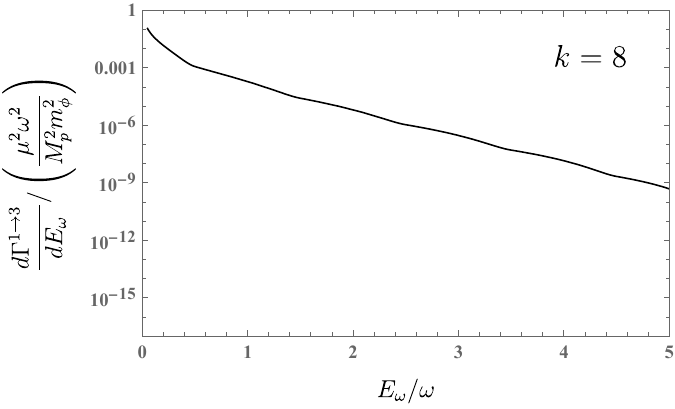}
            \includegraphics[scale=0.65]{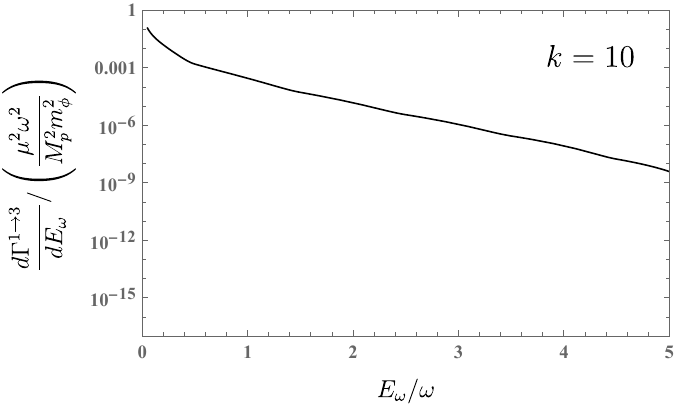}
		\caption{The differential decay rate of the three-body decay for the bosonic decay in the classical field picture. These four panels correspond to $k=4$, $k=6$, $k=8$ and $k=10$, respectively.}
		\label{fig:Differential decay}
\end{figure} 

Although the particle picture used in the literature does not give the same result
as in the classical field picture, the discrepancy can be dissolved if the particle
picture is generalized to the one introduced by Eq.~(\ref{eq:int3.4}) where coherent oscillations
of inflaton is regarded as a collection of an infinite tower of massive particles labeled with 
integers $n$.
Actually, it is straightforward to rewrite the decay rate (\ref{eq:int3.7}) as
\begin{equation}
\frac{d\Gamma_\phi^{1\to 3}}{dE_\omega}=\sum_{n=1}^{\infty}b_n \frac{d\Gamma^{1\to 3}_{\phi_n}}{dE_\omega},
\end{equation}
where $\frac{d\Gamma^{1\to 3}_{\phi_n}}{dE_\omega}$ is the three-body decay rate of a massive 
particle having a mass $n\omega$ computed based on the standard QFT,
\begin{equation}
\frac{d\Gamma_\phi^{1\to 3}}{dE_\omega} = \frac{1}{32\,\pi^3} \left(\frac{\mu}{M_P}\right)^2 \frac{(1 - 2 x_n)^2}{4 x_n}\,, ~~~~~\quad 0<x_n<\frac12.
\end{equation}
Remarkably, the same weight $b_n$ as in Eq.~(\ref{eq:int3.4}) appears in the above decay rate. 
Thus, the graviton bremsstrahlung can be regarded as a process of decays of massive particles 
for which particles labeled with an integer $n$ have mass $n\omega$, constitute a fraction $b_n$,
and decay with the rate $\frac{d\Gamma^{1\to 3}_{\phi_n}}{dE_\omega}$.
Nevertheless, in order to make a comparison between our results and those in the literature,
in what follows, we will retain the original meaning of the particle picture. 


Due to the contributions from higher oscillation modes, the differential decay rate
in the classical field picture shows multiple peaks where peaks are located at 
$\frac{n\omega}{2}$, as it is evident from Fig.~\ref{fig:Differential decay}.
As we have already shown, the differential decay rate in the particle picture does not coincide
with that in the classical field picture.
In \ref{results:boson}, we will investigate the differences between the GW spectra 
computed using the former and the latter pictures,
thereby highlighting the extent of the approximations in the particle picture.

\subsection{Fermionic reheating}
We next consider the fermionic decay, for which the two-body decay rate is given by \cite{Ichikawa:2008ne, Garcia:2020wiy}
\begin{equation}
\Gamma_\phi^{1\to2}(a)=\frac{y_{\text{eff}}^2}{8\pi}m_\phi(a).
\end{equation}
Here $y_{\text{eff}}$ is the effective coupling constant for the fermionic 
decay~\cite{Garcia:2020wiy};
\begin{equation}
y_{\text{eff}}^2(k)=(k+2)(k-1) \left(\frac{\omega}{m_{\phi}}\right)^3 \left[ \sum_{n=1}^{\infty} n^3 |\mathcal{P}_n|^2  \right]y^2,
\end{equation}
which is defined in the same way as $\mu_{\rm eff}$.
Similar to the bosonic case, the two-body decay rate above can be rewritten as 
\begin{equation}
\label{eq:int3.15}
\Gamma_\phi^{1\to2}(a)=\sum_{n=1}^{\infty}b_n\Gamma_{\phi_n}^{1\to2}(a),
\end{equation}
where $\Gamma_{\phi_n}^{1\to2}(a)$ defined by
\begin{equation}
\Gamma_{\phi_n}^{1\to2}(a)=\frac{y^2}{8\pi}n\omega(a)
\end{equation}
is the decay rate of a massive particle with mass $n\omega$.
This result again supports the view that coherent oscillations of inflaton
is a collection of an infinite tower of massive particles where the particles 
labeled with $n$ have a mass $n\omega$, constitute a fraction $b_n$,
and decay into fermions $f$ with the standard decay rate $\Gamma_{\phi_n}^{1\to2}(a)$.

Like the bosonic case, we can calculate the three-body decay rate in the particle picture by
assigning $y_{\rm eff}$ for the vertices of the Feynman diagrams.
The result is given by \cite{Nakayama_2019, Barman:2023rpg}
\begin{align} \label{eq:dGdE_fer}
	\frac{d\Gamma_\phi^{1\to 3}}{d\Eom} = \frac{\yeff^2}{64\,\pi^3} \left(\frac{m_\phi}{M_P}\right)^2\Bigg[&\frac{1 - 2 x}{x} \left(2 x(x - 1) + 1\right) \Bigg]\,,\quad 0<x<\frac12
	\end{align}
Here $x=\frac{E_\omega}{m_\phi}\,$. 
Derivation of the differential decay rate in the classical field picture is given in Appendix~\ref{Sec:Decay Rate}.
The final result is given by
 \begin{equation}\label{eq:int3.17}
\frac{d\Gamma_\phi^{1\to 3}}{dE_\omega}=\frac{y^2}{64\,\pi^3} \left(\frac{m_\phi}{M_P}\right)^2(k+2)(k-1)\left(\frac{\omega}{m_{\phi}}\right)^4\left[ \sum_{n=1}^{\infty} n^4 |\mathcal{P}_n|^2 \frac{1 - 2 x_n}{x_n} \left(2 x_n(x_n - 1) + 1\right) \right]
,\quad 0<x_n<\frac12.
\end{equation}
In the same way as in the bosonic case, this may be re-expressed as
\begin{equation}
\frac{d\Gamma_\phi^{1\to 3}}{dE_\omega}=\sum_{n=1}^{\infty}b_n \frac{d\Gamma^{1\to 3}_{\phi_n}}{dE_\omega},
\end{equation}
where $\frac{d\Gamma^{1\to 3}_{\phi_n}}{dE_\omega}$ is the three-body decay rate of a particle with mass $n\omega$
computed by the standard QFT,
\begin{align} 
	\frac{d\Gamma_{\phi_n}^{1\to 3}}{d\Eom} = \frac{y^2}{64\,\pi^3} \left(\frac{n\omega}{M_P}\right)^2\Bigg[&\frac{1 - 2 x_n}{x} \left(2 x_n(x_n - 1) + 1\right) \Bigg]. 
 \quad 0<x_n<\frac12
	\end{align}
In \cite{Bernal:2023wus}, only the first-order oscillation mode $(n=1)$ in
Eq.~(\ref{eq:int3.17}) was retained. 
Our analysis accounts for the influence of higher-order oscillations.
In the numerical calculations, we take the summation up to the tenth order, 
where the results exhibit good convergence.

\section{Gravitational Wave Spectrum}
Gravitational waves generated by the graviton bremsstrahlung contribute to the 
stochastic GW background of the present universe.
The spectrum of such GW background can be obtained by solving the following
evolution equations for $\rho_\phi, \rho_R$ and $\rho_{\rm GW}$;
\begin{align}
&\dot{\rho}_\phi+3H(1+w_\phi)\rho_\phi=-(1+w_\phi)(\Gamma_\phi^{1\to2}+\Gamma_\phi^{1\to3})\rho_\phi, \label{evo-rho-phi}\\
&\dot{\rho}_\text{R}+4H\rho_\text{R}=(1+w_\phi)\Gamma_\phi^{1\to2}\rho_\phi+(1+w_\phi)\int\frac{d\Gamma_\phi^{1\to 3}}{d\Eom}\frac{m_\phi-E_\omega}{m_\phi}\rho_\phi dE_\omega, \label{evo-rho-R}\\
&\frac{d}{dt}\left(\frac{d\rho_\text{GW}}{dE_\omega}\right)+4H\frac{d\rho_\text{GW}}{dE_\omega}=\frac{2k}{2+k}\frac{\rho_\phi}{m_\phi}\frac{d\Gamma_\phi^{1\to 3}}{dE_\omega}\times E_\omega. \label{evo-rho-GW}
\end{align}
Here $\rho_{\rm GW}$ is not included in $\rho_{\rm R}$ and is considered separately.
As it is evident from (\ref{eq:int3.7}) and (\ref{eq:int3.17}),
for both the cases of the bosonic decay and the fermionic decay,
$d\Gamma_\phi^{1\to 3}/dE_\omega$ diverges like $1/E_\omega$ when $E_\omega\to0$, 
which is known as infrared divergence. 
Therefore $\Gamma_\phi^{1\to3}$ in (\ref{evo-rho-phi}) and $\int d\Gamma_\phi^{1\to 3}$ in (\ref{evo-rho-R}) exhibit logarithmic infrared divergence. 
This divergence should, in principle, be resolved by incorporating higher-order diagrams, which are beyond the scope of our analysis \cite{Addazi:2019mjh}. Instead, following the approach in \cite{Nakayama_2019}, we introduce a low-energy cutoff $\Lambda$ to regularize the divergence, 
setting $\Lambda=10^{-7}m_\phi$. 
As the divergence is only logarithmic, our results remain largely insensitive to the specific value of $\Lambda$, provided it is not exceedingly small. 
\footnote{More specifically, the cutoff $\Lambda$ does not affect the final results as long
as it is larger than $e^{-M_p^2/m_\phi^2}m_\phi$ which is exponentially suppressed
compared to $m_\phi$. 
The value of \(\Lambda\) we adopt is much larger than this value, 
and therefore, it will not affect the final results.}
Under this prescription, we find that in (\ref{evo-rho-phi}) and (\ref{evo-rho-R}), the relative magnitude of the three-body decay terms is highly suppressed($\lesssim {\cal O}(10^{-20})$) compared to the two-body decay terms. 
Therefore, we will ignore them in the subsequent calculations.
We numerically solve these equations to determine the time evolution of the dynamical quantities and finally the GW spectrum at the present time. 
The initial conditions that we impose at the time of the end of inflation are 
$\rho_\text{R}=\rho_{\rm GW}=0$. 
The initial value of $\rho_\phi$ can be obtained from the value of the inflaton 
at the end of inflation $\phi_\text{end}$ given by \cite{Garcia:2020wiy}
\begin{equation}
\phi_\text{end}=\sqrt{\frac{3}{8}}M_p \ln\left[\frac12+\frac{k}3\left(k+\sqrt{k^2+3}\right)\right].
\end{equation}
Its end is defined as the condition $\ddot{a}=0$, which can be shown to be equivalent to $\dot{\phi}_{\text {end }}^2=V(\phi_{\text {end }})$ \cite{Ellis:2015pla}. We can determine $\rho_\phi$ at this time to be 
\begin{equation}
\rho_\phi=\frac{3}{2}\lambda M_p^4 \left(\frac{\phi_\text{end}}{M_p}\right)^k.
\end{equation}
For these initial conditions and the differential equations, there are three free 
parameters; the coupling constant (either $\mu$ or $y$), $\lambda$ in the potential, 
and the exponent $k$.  
In the numerical calculations, we fix the energy scale by setting $\lambda=10^{-11}$
and study how the resultant GW spectrum varies against the change of $k$ ($k=2$ to $k=10$) and the coupling constants. 
We take into account the red-shift of the graviton's energy to translate the GW
spectrum from the time of reheating to that in the present universe.
The GW spectrum at the present time is then defined as
\begin{equation}
\Omega^{\text{0}}_\text{GW}=\frac{1}{\rho^\text{0}_\text{c}}\left.\frac{d\rho_\text{GW}}{d\ln E_\omega}\right|_{a_0},
\end{equation}
where $\rho^\text{0}_\text{c}$ is the critical energy density today.
The figures given in the next section are plotted in terms of this quantity.

\section{Numerical Results}
\label{sec:numerical-results}

\begin{figure}[t]
		\centering
		\includegraphics[scale=0.37]{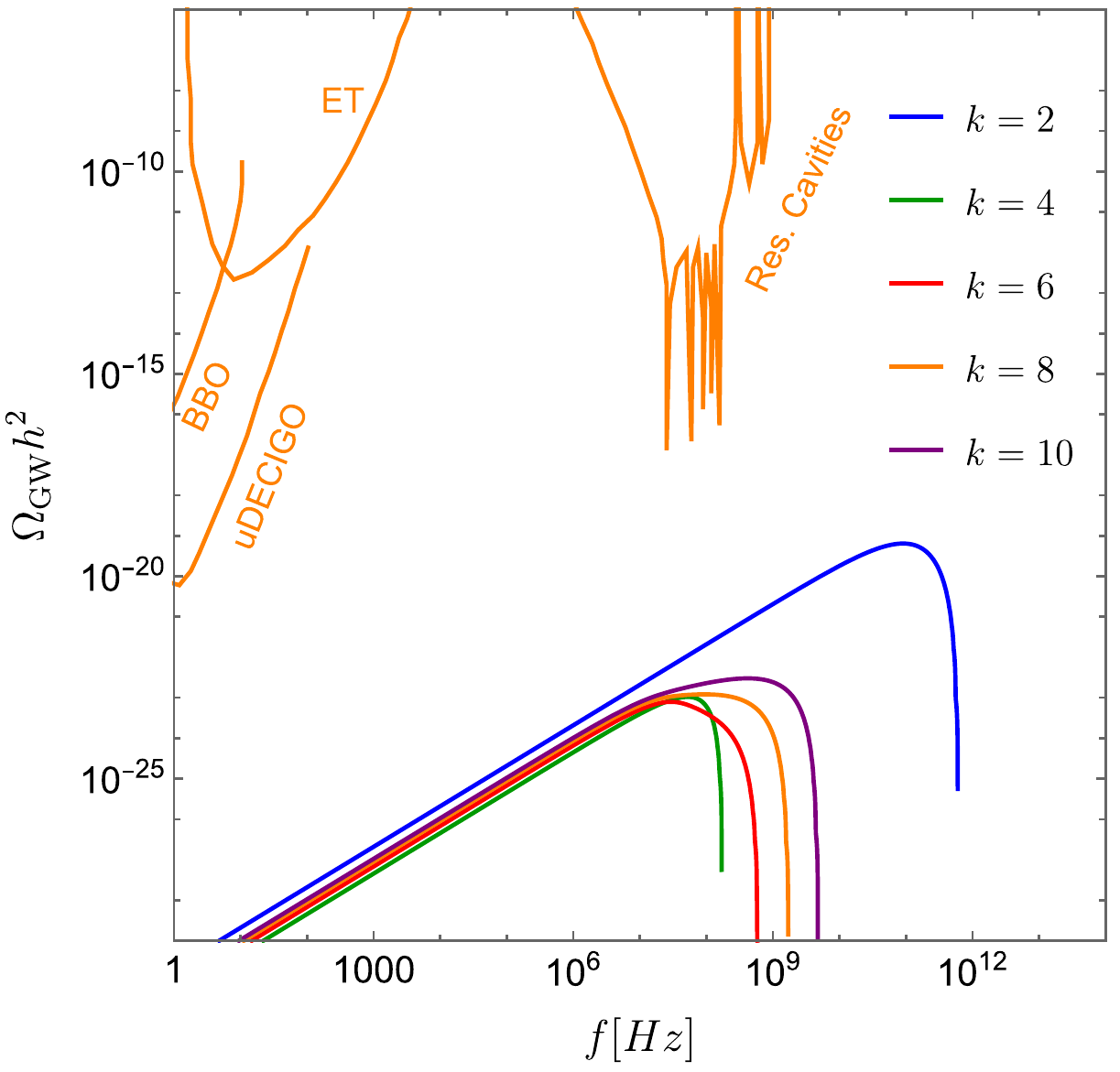}
		\includegraphics[scale=0.37]{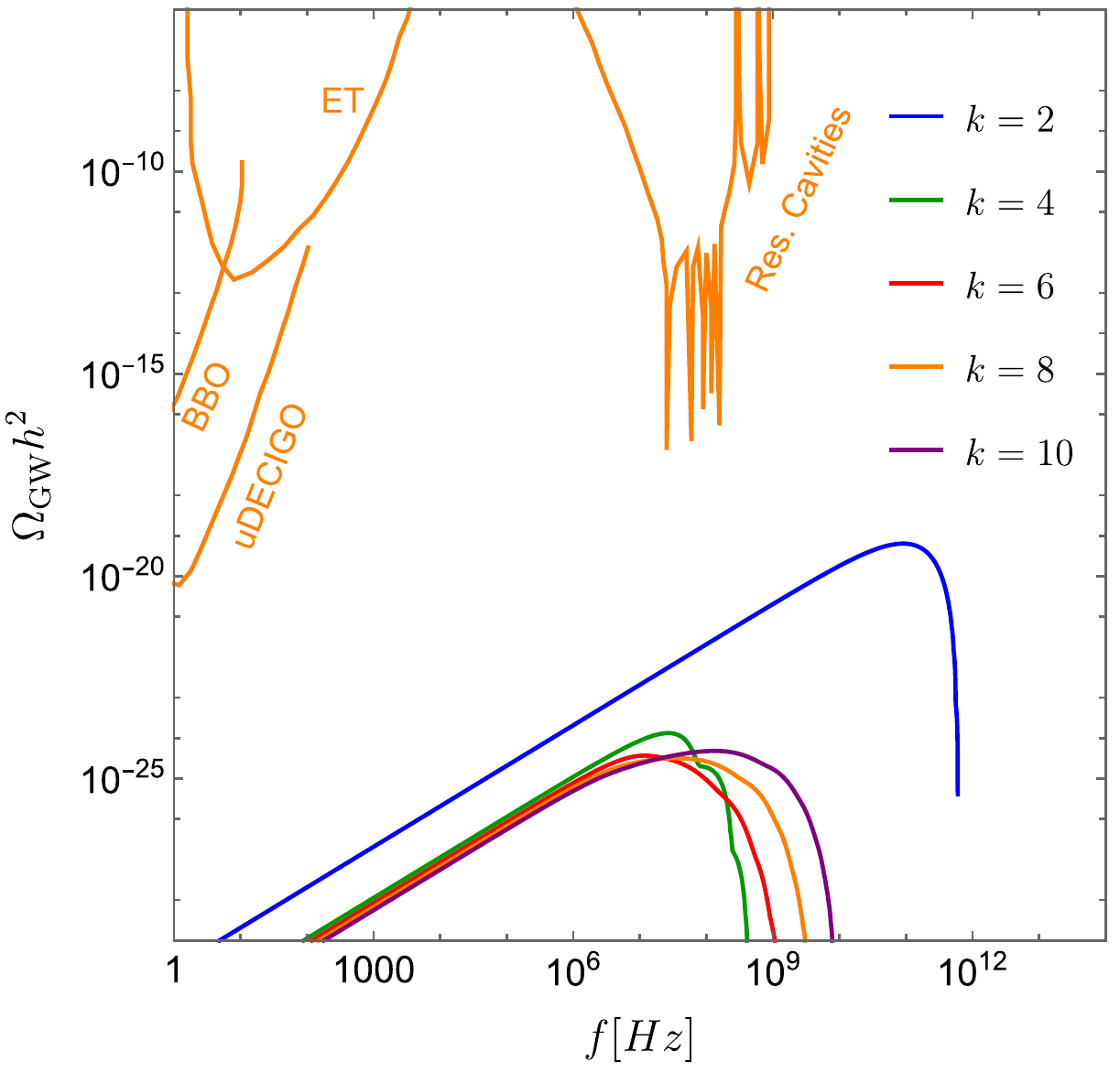}
		\caption{GW spectra are shown for the particle picture (left panel) and the 
  classical field picture (right panel) in the case where the inflaton decays to bosons. 
  The exponent $k$ is from $k=2$ to $k=10$, $\lambda=10^{-11}$, and $\mu=10^{10}\,\text{GeV}$. The solid orange lines represent the sensitivities of future gravitational wave detectors.}
		\label{fig:Omegaboson}
\end{figure} 
In this section, we show the present-day spectrum of the stochastic GWs originating from the graviton bremsstrahlung based on the classical field picture by numerically
solving the evolution equations (\ref{evo-rho-phi}), (\ref{evo-rho-R}),
and (\ref{evo-rho-GW}).
For comparison, we also show the results based on the particle picture.
For the GW spectrum, we fix the coupling constant first and investigate how the results depend on the value of the 
exponent $k$.

\subsection{Bosonic reheating}
\label{results:boson}

For the case of the decay to bosons, we fix $\mu$ to $\mu=10^{10}\,\text{GeV}$. 
The left/(right) panel of Fig.~\ref{fig:Omegaboson} shows the GW spectrum calculated based
on the particle picture/(classical field picture).
In both cases, we find that both the amplitude of the GW spectrum 
and the peak frequency becomes the largest for $k=2$. 
For the other values of $k$, the amplitude of the GW spectrum decreases slightly 
and the maximum frequency increases as $k$ is increased. 

For the cases based on the classical field picture, we see an obvious multi-peak shape 
in the case $k=4$, just as we expect from the multi-peak structure of the 
differential decay rate derived in Sec.~\ref{Sec:Bremsstrahlung}.
This feature becomes less pronounced as $k$ is increased. 
In particular, in the cases $k=8$ and $k=10$, multiple peaks are significantly suppressed and hard to recognize visually. 
In the case $k=4$, the coefficients ${\cal P}_n$ decay quickly as $n$ is increased, which explains the clear feature of the multiple peaks 
in the GW spectrum. 
On the other hand, for the cases $k=8$ or $k=10$, such a large hierarchy among ${\cal P}_n$ does not exist for the first several oscillation modes,
and contributions from different oscillation modes tend to average and erase the multi-peak
structure. 

In the figure, in addition to the GW spectrum, the expected sensitivity curves of
the following future GW detectors are also included: 
Big Bang Observer (BBO)~\cite{Crowder_2005,Corbin_2006}, the ultimate Deci-hertz Interferometer Gravitational Wave Observatory (uDECIGO)~\cite{Seto_2001,PhysRevD.73.064006}, the Einstein Telescope (ET)~\cite{Punturo:2010zz,Hild_2011,Sathyaprakash_2012,Maggiore_2020}, 
and the Resonant Cavities~\cite{Berlin_2022,Herman_2023} which target high-frequency GWs and 
is still in the conceptual stage. 
It is found that the signals are below the sensitivity curves.
Thus, detectors with better sensitivity are needed to reach the signals.

 \begin{figure}[h]
		\centering
		\includegraphics[scale=0.6]{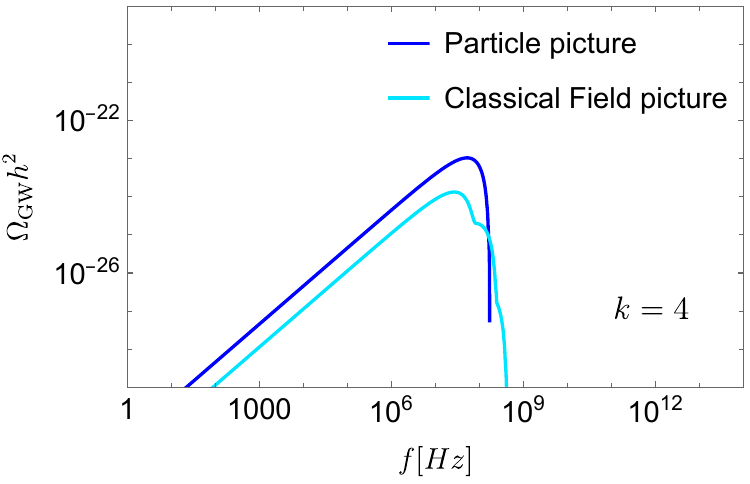}
		\includegraphics[scale=0.6]{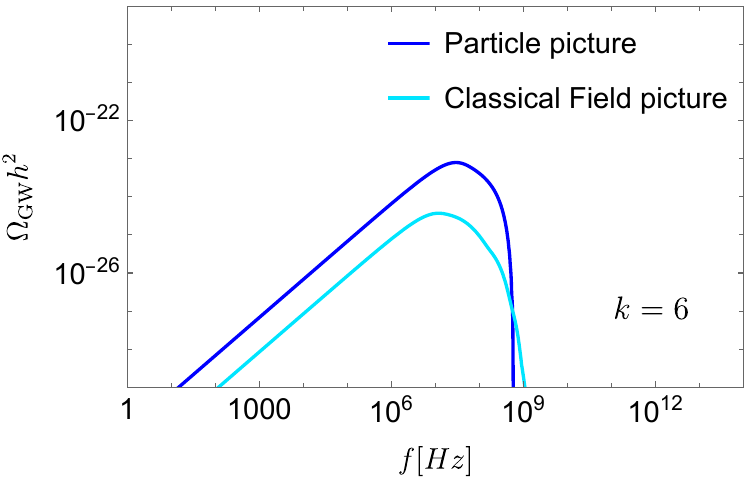}
            \includegraphics[scale=0.6]{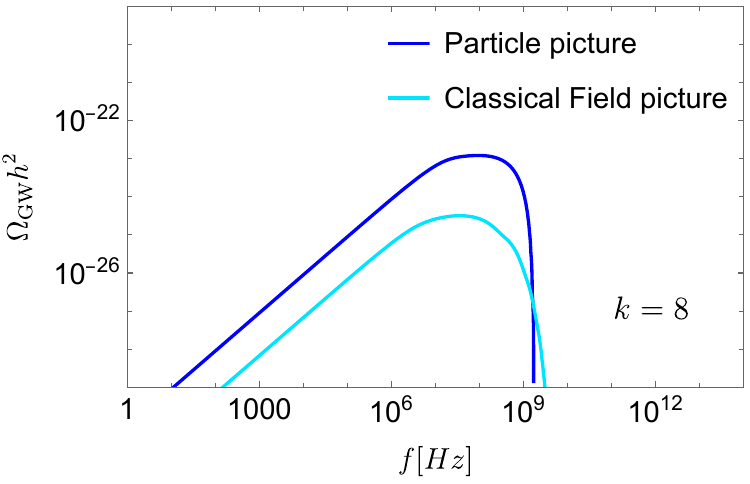}
            \includegraphics[scale=0.6]{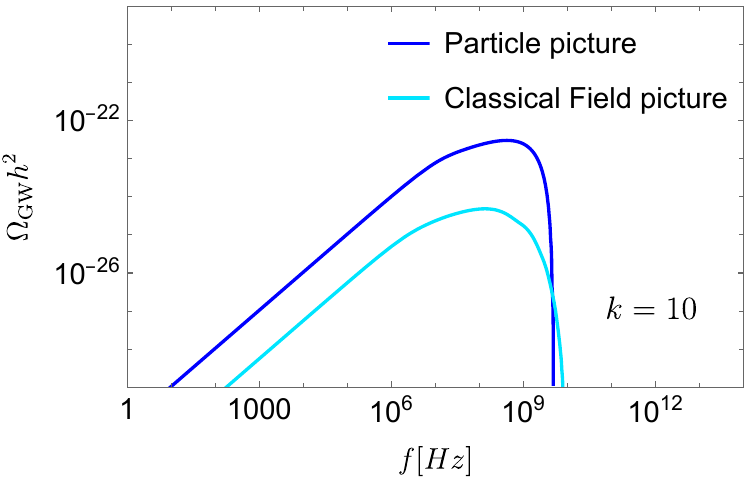}
		\caption{GW spectra calculated based on the particle picture (dark blue) and the classical field picture (light blue) in the case where inflaton decays to bosons. These four panels correspond to $k=4, k=6, k=8$ and $k=10$, respectively.}
		\label{fig:Omegabosoncompare}
\end{figure} 

We next compare the GW spectrum calculated based on the particle picture and 
the classical field picture. 
The four panels in Fig.~\ref{fig:Omegabosoncompare} are comparison figures 
between the two pictures for $k=4, k=6, k=8$, and $k=10$. 
The case $k=2$ is not included because the inflaton has only one oscillation mode and 
the two pictures give the same results. 
From the figures, we observe that, except for high-frequency region, 
the amplitude of the GW spectrum in the particle picture 
exceeds that in the classical field picture for any value of $k$.
The difference, quantified by the ratio of amplitudes, ranges from ${\cal O}(10)$ to 
${\cal O}(100)$, increasing with higher $k$.
Therefore, the particle picture tends to overestimate the amplitude of the GW spectrum.

As for the scaling of the GW spectrum on the coupling constant, analytical derivation is given in Appendix~\ref{Sec:coupling constant} and here we only give the results. 
First, the peak frequency of GW spectrum scales with the coupling 
constant $\mu$ as
\begin{equation}
f_{\mathrm{GW}, 0} \propto
\begin{cases}
    \mu^{\frac{k-4}{2k-2}} &  (k \leq 6)\,,\\
    \mu^{-\frac{k-4}{6(k-1)}} & (k \geq 8)\,.
\end{cases}
\end{equation}\label{eq:frequency dependence}
The distinction between $k\leq 6$ and $k\geq 8$ arises because of the difference of
epochs when the GWs are dominantly produced.
The height of the GW spectrum at the peak frequency 
depends on the coupling constant $\mu$ as
\begin{equation}
\Omega_{\rm GW}^0 \propto
\begin{cases}
    \mu^{\frac{2(k-2)}{k-1}} &  (k \leq 6)\,,\\
    \mu^{\frac{4k+2}{3k-3}} & (k \geq 8)\,.
\end{cases}
\end{equation}\label{eq:height dependence}

\begin{figure}[t]
		\centering
		\includegraphics[scale=0.6]{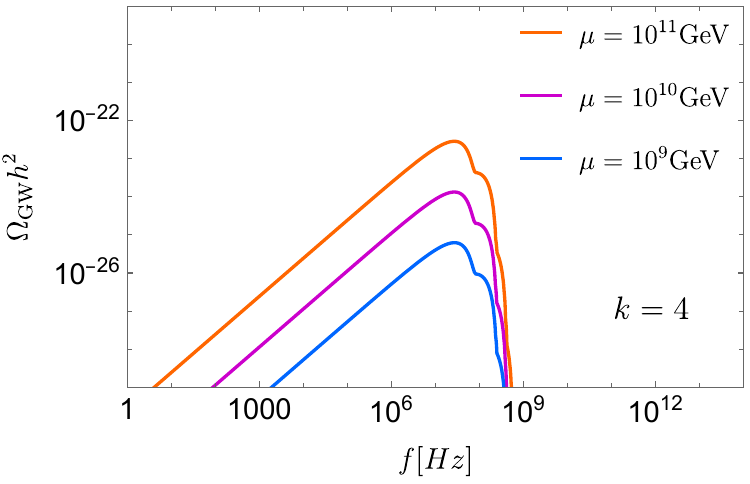}
		\includegraphics[scale=0.6]{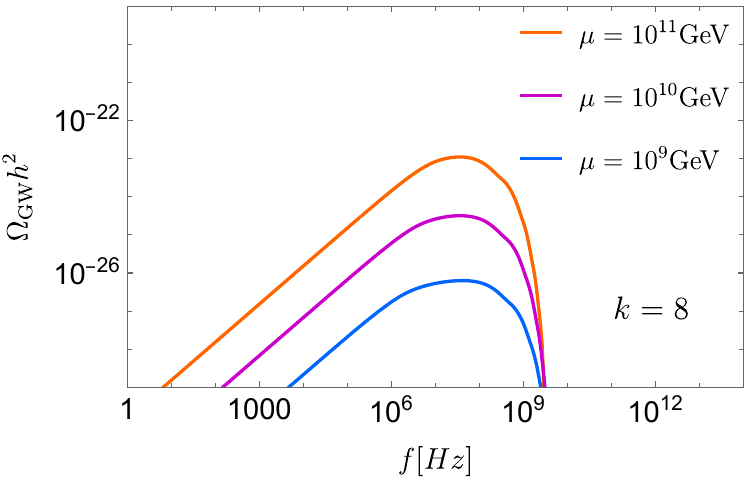}
		\caption{The dependence of GW spectrum on coupling constant $\mu$ in bosonic reheating. The left figure is $k=4$ case and the right figure is $k=8$ case. In both cases, the coupling constants are set to $\mu=10^{11}~\text{GeV}$, $\mu=10^{10}~\text{GeV}$, and $\mu=10^9~\text{GeV}$ respectively.   }
		\label{fig:Boson_dependence}
\end{figure}

For instance, we have
\begin{align}
&f_{\mathrm{GW}, 0} \propto \text{contant},\quad\Omega_{\rm GW} \propto \mu^{\frac{4}{3}}\qquad (k=4) \\
&f_{\mathrm{GW}, 0} \propto \mu^{-\frac{2}{21}},\quad\Omega_{\rm GW} \propto \mu^{\frac{34}{21}}\qquad (k=8)
\end{align}
To illustrate that these analytic scalings are consistent with the results of numerical calculations, 
in Fig.~\ref{fig:Boson_dependence} we show the GW spectra computed based on the classical field picture
for two cases $k=4$ (left panel) and $k=8$ (right panel) for three different values of the
coupling constant $\mu=10^{11}~\text{GeV}$, $\mu=10^{10}~\text{GeV}$, and $\mu=10^9~\text{GeV}$.
We confirm that the numerical results show good consistency with the analytical scalings.

In the same manner as we have derived the scaling of the GW spectrum with the coupling constant, 
the dependence of the GW spectrum on the parameter \(\lambda\) can also be derived. 
First, it can be shown that the peak frequency of GW spectrum scales with the parameter 
$\lambda$ as
\begin{equation}
f_{\mathrm{GW}, 0} \propto
\begin{cases}
    \lambda^{\frac{1}{4k-4}} &  (k \leq 6)\,,\\
    \lambda^{\frac{4k-13}{12(k-1)}} & (k \geq 8)\,.
\end{cases}
\end{equation}\label{eq:frequency lambda dependence}
The height of the GW spectrum at the peak frequency depends on the parameter $\lambda$ as
\begin{equation}
\Omega_{\rm GW}^0 \propto
\begin{cases}
    \lambda^{\frac{1}{k-1}} &  (k \leq 6)\,,\\
    \lambda^{\frac{k-4}{3k-3}} & (k \geq 8)\,.
\end{cases}
\end{equation}\label{eq:height lambda dependence}

Furthermore, in the case of \(k=2\), the inflaton may undergo fragmentation through a process of tachyonic resonance when coupled to a boson via a trilinear interaction \cite{Dufaux:2006ee}. Specifically, the trilinear coupling \(\mu\phi b^2\) contributes a tachyonic squared mass term of the form \(m_b^2 \sim \mu\phi\) to the boson. This squared mass becomes negative during certain phases of \(\phi\)'s oscillation, leading to tachyonic resonance. However, when considering the self-interaction of bosonic fields, represented by the term \(\lambda_b b^4\), a positive squared mass term \(\lambda_b\langle b^2\rangle\) is introduced. This backreaction suppresses the tachyonic effects \cite{Dufaux:2006ee,Bernal:2023wus,xu2024ultrahighfrequencygravitationalwaves}.
In our study, we assume the absence of both the trilinear and self-interactions.

\subsection{Fermionic reheating}

\begin{figure}[t]
		\centering
		\includegraphics[scale=0.37]{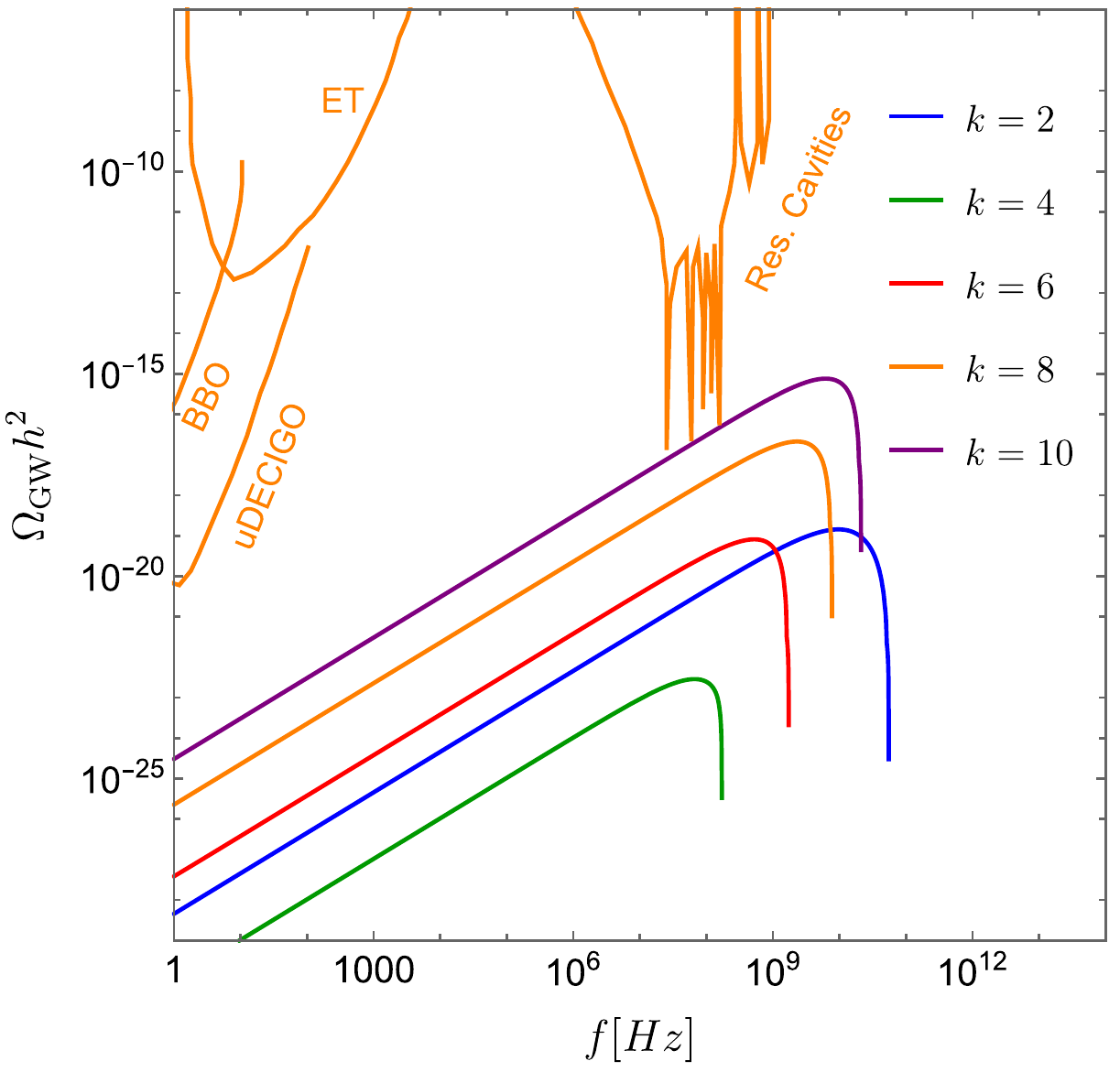}
		\includegraphics[scale=0.37]{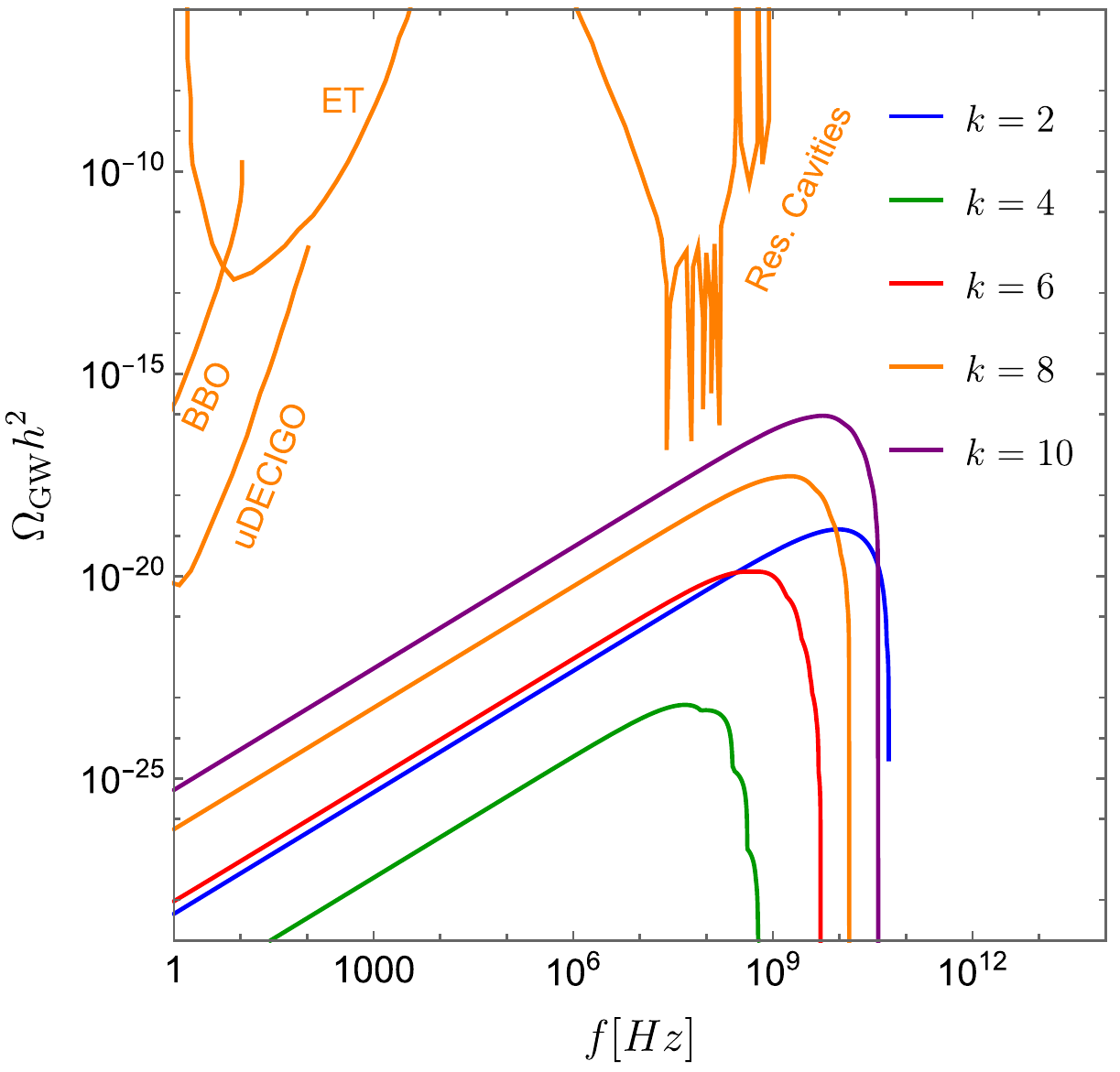}
		\caption{GW spectrum based on the particle picture (left panel) and the 
  classical field picture (right panel) in the case where inflaton decays to fermions. 
  The exponent $k$ is from $k=2$ to $k=10$, $\lambda=10^{-11}$ and $y=0.01$. Solid orange lines represent the sensitivities of future gravitational wave detectors.}
		\label{fig:Omegafermion}
\end{figure} 

For the case of the decay to fermions, we fix $y$ to $y=0.01$. 
The left/(right) panel of Fig.~\ref{fig:Omegafermion} shows the GW spectrum calculated in the 
particle picture/(the classical field picture), respectively. 
The multi-peak structure seen in the right panel (especially for $k=4$) is again caused by the
higher-order oscillation modes of inflaton.
Contrary to the decays to bosons, the GW spectrum is greatly enhanced as $k$ increases except for $k=2$. 
As a result, in the extreme case $k=10$, the GW signal is close to the sensitivities of resonant cavity detectors,
suggesting the possibility of detecting GWs from graviton bremsstrahlung in the future.

 \begin{figure}[t]
		\centering
		\includegraphics[scale=0.6]{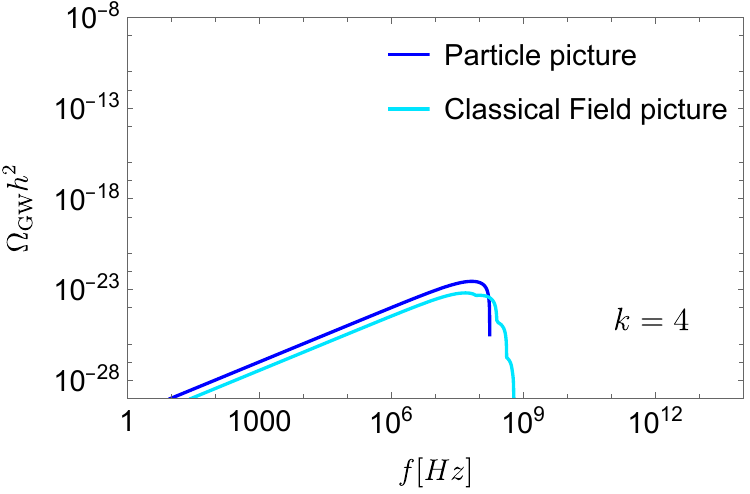}
		\includegraphics[scale=0.6]{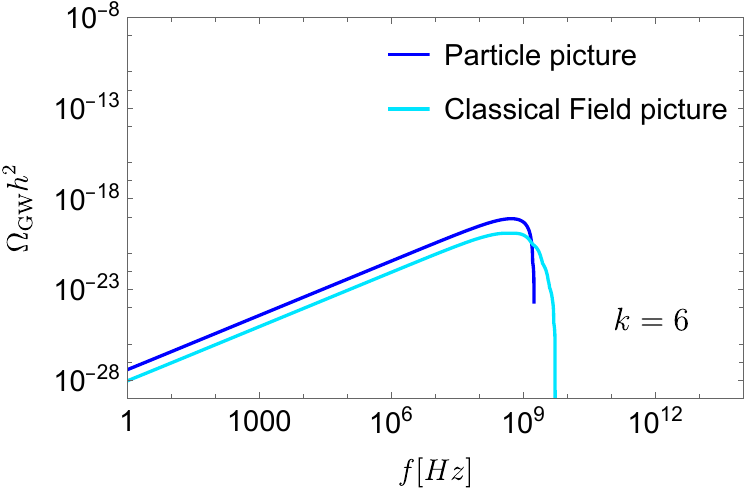}
            \includegraphics[scale=0.6]{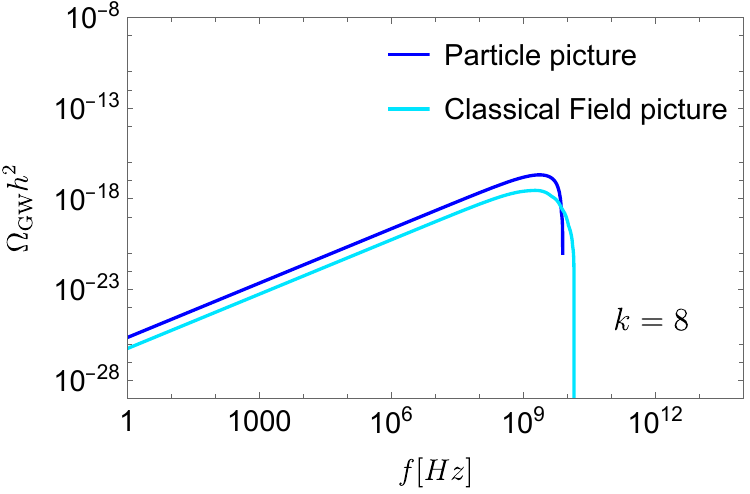}
            \includegraphics[scale=0.6]{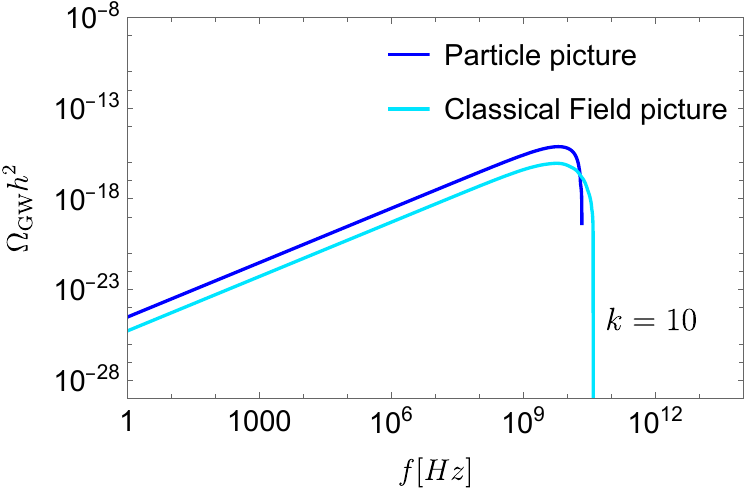}
		\caption{GW spectrum calculated from QFT method(dark blue) and calculated from classical field method(light blue). These four figures correspond to k=4, k=6, k=8 and k=10. Solid orange lines represent the sensitivities of future gravitational wave detectors.}
		\label{fig:Omegafermioncompare}
\end{figure} 

We then compare the GW spectra calculated based on the particle picture and 
the classical field picture for four different cases $k=4, 6, 8, 10$ in Fig.~\ref{fig:Omegafermioncompare}. 
As it is found in the case of the bosonic decay,
we find that GW spectra calculated in the particle picture always overestimate 
the amplitude compared to the classical field picture, except for the high-frequency part.

In the appendix~\ref{Sec:coupling constant}, we provide a full derivation of the analytic scaling of the peak frequency 
and amplitude of 
GW spectrum against 
the variation of the coupling constant $y$.
The results are given by
\begin{equation}
f_{\mathrm{GW}, 0} \propto
\begin{cases}
    y^{-1} &  (k = 2)\,,\\
    y^{\frac{k-4}{6}} &  (4 \leq k \leq 6)\,,\\
    y^{\frac{1}{2}} & (k \geq 8)\,.
\end{cases}
\end{equation}\label{eq:frequency dependenceF}
and
\begin{equation}
\Omega_{\rm GW} \propto
\begin{cases}
    \text{contant} &  (k = 2)\,,\\
    y^{\frac{-2k+14}{3}} &  (4 \leq k \leq 6)\,,\\
    \text{contant} & (k \geq 8)\,.
\end{cases}
\end{equation}\label{eq:height dependenceF}

\begin{figure}[t]
		\centering
		\includegraphics[scale=0.6]{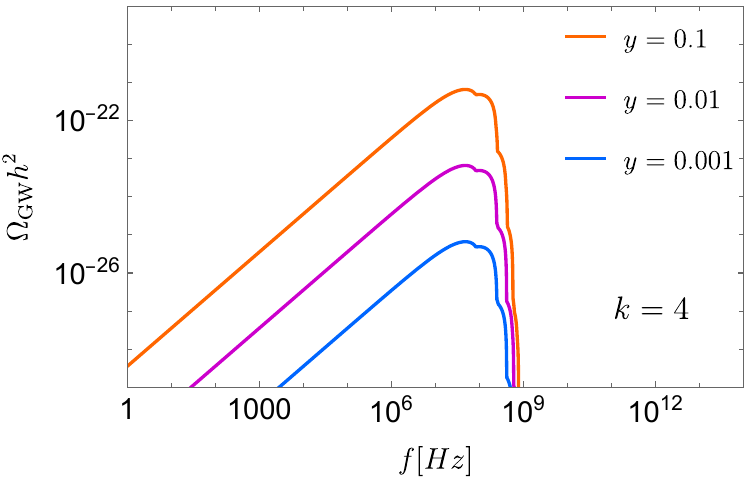}
		\includegraphics[scale=0.6]{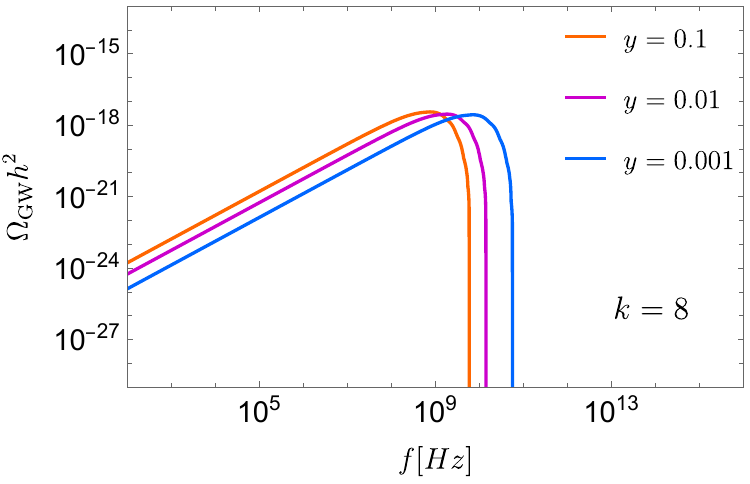}
		\caption{The dependence of GW spectrum on coupling constant $y$ in fermionic reheating. The left figure is $k=4$ case and the right figure is $k=8$ case. In both cases, the coupling constants are set to  $y=0.1$, $y=0.01$, and $y=0.001$ respectively.   }
		\label{fig:Fermion_dependence}
\end{figure}

As in the bosonic case, to illustrate that these analytic scalings are consistent with the results of numerical calculations, two cases $k=4, 8$ are considered as examples. 
In both cases, the coupling constants are set to $y=0.1$, $y=0.01$, and $y=0.001$ respectively. 
Using the above scalings, we have
\begin{align}
&f_{\mathrm{GW}, 0} \propto \text{contant},\quad\Omega_{GW} \propto y^{-2}\qquad (k=4) \\
&f_{\mathrm{GW}, 0} \propto y^{\frac{1}{2}},\quad\Omega_{GW} \propto \text{contant}\qquad (k=8)
\end{align}
Fig.~\ref{fig:Fermion_dependence} shows the GW spectra based on the numerical calculations under 
the classical field picture. 
It is confirmed that the analytical scalings are consistent with the numerical results.

The dependence of the GW spectrum on the potential parameter \(\lambda\) can also be derived as bosonic case. First, the peak frequency of GW spectrum scales with the parameter $\lambda$
\begin{equation}
f_{\mathrm{GW}, 0} \propto \lambda^{\frac{1}{4}}
\end{equation}\label{eq:Fermion frequency lambda dependence}
The height of the GW spectrum at the peak frequency depends on the parameter $\lambda$ as
\begin{equation}
\Omega_{\rm GW}^0 \propto \lambda
\end{equation}\label{eq:Fermion height lambda dependence}

\subsection{Discussions}
The feature of multi-peaks arises from the superposition of different oscillation modes of the inflaton, making it a unique characteristic of gravitons produced by the decay of the inflaton oscillating 
around the potential with $k\geq 4$. 
For the other bremsstrahlung mechanisms generating gravitons such as the one 
from the decay of massive particles or particles produced by primordial black hole evaporation, the shape of the gravitational wave spectrum is similar, 
but they do not exhibit the multi-peaks feature~\cite{Choi_2024}. 
This suggests an interesting possibility that observational confirmation of the presence of the multi-peaks provides the strong evidence that the gravitons originate from the decay of the inflaton with the potential with $k\geq 4$.

For cases where \(k > 2\), it is known that the inflaton condensate undergoes fragmentation due to self-interactions, resulting in inhomogeneities in the inflaton \cite{Lozanov:2016hid,Lozanov:2017hjm,Garcia:2023eol,Garcia:2023dyf}. 
Previous studies have shown that this fragmentation typically occurs more than 5 e-folds 
after the end of inflation. 
As discussed in Appendix C, for \(k > 2\), GWs from graviton bremsstrahlung 
are dominantly produced near the beginning of the oscillatory phase, 
except in the cases of bosonic reheating with \(k = 4\) and \(k = 6\). 
Therefore, if such bosonic decay is not the main channel,  
the final GW spectrum derived in this paper remains unchanged,
even when accounting for the effects of fragmentation. 
For bosonic reheating with \(k = 4\) and \(k = 6\), however,
GWs are dominantly generated around the time of reheating.
Even in such scenarios, if inflaton fragmentation occurs after reheating, 
its impact on the GW spectrum will be not significant. 
According to \cite{Lozanov:2017hjm}, self-fragmentation is completed by the inflaton quanta produced within the first narrow instability band described by the Floquet theorem. 
This provides an estimate for the time at which backreaction from fluctuations becomes significant.
Using the results of this paper and imposing the condition that the reheating time 
is less than the backreaction time, a lower bound on the coupling constant $\mu$ can be derived. 
For instance, assuming $\lambda=10^{-11}$, we find that the bound is given by
\(\mu > 2.87 \times 10^{10} \, \text{GeV}\) for \(k = 4\) and 
\(\mu > 1.29 \times 10^{6} \, \text{GeV}\) for \(k = 6\), respectively.
If $\mu$ falls below these thresholds, the effects of fragmentation on the final GW spectrum
will become significant.
Computing the GW spectra including the effects of fragmentation is beyond the scope of this paper.

Moreover, in this study, gravitational reheating is assumed to be negligible. Previous research on gravitational reheating has demonstrated that for \( w_\phi \gtrsim 0.65 \)~\cite{Clery:2021bwz,Haque:2022kez,Haque:2023zhb,Co:2022bgh,Barman:2022qgt}, corresponding to \( k > 9 \), gravitational reheating can independently reheat the universe and satisfy the temperature constraints imposed by BBN. As \( k \) increases further, gravitational reheating becomes a non-negligible effect. Therefore, this study does not consider potentials with \( k > 10 \).

\section{Conclusion}
\label{sec:conclusion}

Graviton bremsstrahlung, the production of gravitons accompanied with the decay of inflaton, is an inevitable physical process and offers a potential window into how inflation terminated and 
transitioned to reheating.
In light of this situation, we investigated gravitational waves generated by graviton bremsstrahlung
in scenarios where the inflaton oscillates in a polynomial potential, $V(\phi) \simeq \phi^k$, 
allowing for $k>2$. 
In the literature, two different methods are employed to compute the decay rate in such scenarios.
The first, which we refer to as the {\it particle picture}, treats the inflaton as a large number 
of rest particles, and the decay rate is computed using 
standard perturbation theory in quantum field theory.
The second, which we term the {\it classical field picture}, treats the inflaton as an external classical field that imparts time variation to the coupling constant of interaction terms 
between the inflaton and light fields.
The decay rate is then computed using this time-dependent coupling, with the vacuum as the initial state.

While the differential decay rate of graviton bremsstrahlung is available in
the literature for the particle picture, it has not been derived for the classical field picture. 
One of the key contributions of this paper is deriving the differential decay rate in the classical
field picture.

Our results reveal that the differential decay rates calculated using the two methods differ significantly for $k\ge 4$.
Specifically, while the particle picture restricts the production of gravitons with energies 
larger than half the inflaton mass,
the classical field picture allows for the production of gravitons with arbitrarily large energies.
This discrepancy arises because, for $k\ge 4$, the inflaton's oscillations are no longer 
described by a single harmonic mode but instead consist of infinitely many harmonic modes 
with different frequencies.
However, this discrepancy can be resolved by generalizing the particle picture:
if we treat the coherent oscillations of the inflaton as a collection of an infinite tower 
of massive particles, each having the same coupling to daughter particles,
the two methods can be reconciled.
In this view, the graviton bremsstrahlung can be understood as the decay
of these massive particles.

Beyond this qualitative difference, 
we also performed a numerical analysis to quantify the deviation between the two methods. 
Our results show that the particle picture overestimates the amplitude of the gravitational-wave spectrum at the peak frequency by a factor of ${\cal O}(10)$ to ${\cal O}(100)$, depending
on the power index $k$.
This suggests that the particle picture should not be used to accurately determine
the gravitational-wave spectrum for $k\ge 4$.

Moreover, due to the superposition of different inflaton oscillation modes, 
the gravitational wave power spectrum exhibits distinct multi-peak features. 
This unique signal can help distinguish between various inflaton potentials during the oscillation period, 
and also distinguish graviton bremsstrahlung in this context from other scenarios.

In conclusion, our study strengthens the theoretical foundation for using high-frequency gravitational waves to probe the post-inflationary oscillation phase.
Future observations of such waves from the early universe could provide
deeper insights into the physics of reheating.

\section*{Acknowledgments}
We would like to thank Drazen Glavan for giving us useful comments. This work was supported by the JSPS KAKENHI Grant Number JP23K03411 (TS). Y.J. was supported by JST SPRING, Japan Grant Number JPMJSP2180.

\appendix

\section{Weight factor $b_n$}
\label{Sec:bn}
Using~\eqref{eq:int2.16}, the energy density and the pressure stored in the inflaton field can be rewritten as
\begin{align}
\rho_\phi&\simeq\frac12 \langle\dot{\phi}^2\rangle+\langle V(\phi)\rangle\simeq\frac{k+2}2\langle V(\phi)\rangle=V(\phi_0)\label{eq:intB1}\\
P_\phi&\simeq\frac12 \langle\dot{\phi}^2\rangle-\langle V(\phi)\rangle\simeq \frac{k-2}2\langle V(\phi)\rangle=\frac{k-2}{k+2}V(\phi_0)\label{eq:intB2}
\end{align}
From~\eqref{eq:int2.16} and~\eqref{eq:intB1}, there is a relation, 
\begin{equation}
\langle\dot{\mathcal{P}}^2\rangle=\frac{\langle\dot{\phi(t)}^2\rangle}{\phi_0^2}=\frac{k\langle V(\phi)\rangle}{\phi_0^2}=\frac{2k}{k+2} \frac{\rho_\phi}{\phi_0^2}\label{eq:intB3}
\end{equation}
The effective mass can be rewritten as 
\begin{equation}
m_\phi^2(t)=V''(\phi_0)=k(k-1)\frac{\rho_\phi}{\phi_0^2}\label{eq:intB4}
\end{equation}
Combining \eqref{eq:intB3} and~\eqref{eq:intB4} gives 
\begin{equation}
\langle\dot{\mathcal{P}}^2\rangle=\frac{2k}{k+2}\frac{m_\phi^2}{k(k-1)}\label{eq:intB5}
\end{equation}
Using the expansion of $\mathcal{P}$ given by (\ref{eq:int2.12}), the left side of the above equation 
becomes
\begin{equation}
\label{eq:intB6}
\langle\dot{\mathcal{P}}^2\rangle=2\sum_{n=1}^{\infty}|\mathcal{P}_n|^2n^2\omega^2.
\end{equation}
Combining \eqref{eq:intB5} and~\eqref{eq:intB6}, we obtain a relation
\begin{equation}
\sum_{n=1}^{\infty}(k+2)(k-1)\left(\frac{\omega}{m_\phi}\right)^2n^2|\mathcal{P}_n^2|=1.
\end{equation}
From the definition of $b_n$ (\ref{def-bn}), we finally obtain the relation for the weight factor $b_n$ as
\begin{equation}
\sum_{n=1}^{\infty}b_n=1.
\end{equation}
This proves that $b_n$ is a normalized weight parameter.

 \section{Three-body Differential Decay Rate}
	\label{Sec:Decay Rate}
Here we explain how to calculate the three-body decay rate by treating the inflaton as a classical field. The evolution equation of inflaton energy density in the presence of the decay to radiation is given by
\begin{equation}
\dot{\rho}_\phi+3H(1+w_\phi)\rho_\phi=-\Gamma_\phi(1+w_\phi)\rho_\phi.
\end{equation}
The right-hand side represents the energy transfer per space-time volume, 
\begin{equation}
(1+w_\phi)\Gamma_\phi \rho_\phi\equiv\frac{\Delta E}{\text{Vol}_4}.
\end{equation}
The energy transfer can be calculated by
\begin{equation}
\Delta E=\int\frac{d^3\mathbf{p}_A}{(2\pi)^3 2p_A^0}\frac{d^3\mathbf{p}_B}{(2\pi)^3 2p_B^0}\frac{d^3\mathbf{p}_C}{(2\pi)^3 2p_C^0}(p_A^0+p_B^0+p_C^0)|\langle f|i\int d^4x \mathcal{L}_\text{int}|0\rangle|^2, 
\end{equation}
where $p_A$, $p_B$, $p_C$ are the momenta of three particles in the final state, 
and $\mathcal{L}_\text{int}$ is the interaction Lagrangian. 
Assuming ${\cal L}_{\rm int}$ is linear in $\phi$, the transition amplitude may be written as
\begin{equation}
|\langle f|i\int d^4x \mathcal{L}_\text{int}|0\rangle|^2=\text{Vol}_4\sum_{n=-\infty}^{\infty}|\mathcal{M}_n|^2(2 \pi)^4\delta^4(p_n-p_A-p_B-p_C).
\end{equation}
Here $\mathcal{M}_n$ denote the contribution to the transition amplitude from
$n$-th oscillation mode of inflaton.  
We should replace the inflaton field $\phi$ as
\begin{equation}
\phi(t)=\phi_0(t)\sum_{n=-\infty}^{\infty}\mathcal{P}_n e^{-in\omega t}
\end{equation}
Because $\phi_0 \mathcal{P}_n$ is the coefficient term of the amplitude, it can be factored out, and the remaining part can be defined as $\mathcal{M}'_n$. 
\begin{equation}
\left|\mathcal{M}_n\right|^2=\phi_0^2\left|\mathcal{P}_n\right|^2\left|\mathcal{M}'_n\right|^2
\end{equation}
The decay rate is that
\begin{equation}\label{eq:A5}
\Gamma_\phi=\frac{1}{(1+w_\phi)\rho_\phi}\sum_{n=1}^{\infty}\phi_0^2|\mathcal{P}_n|^2n\omega\int dE_pdE_\omega |\mathcal{M}'_n|^2
\end{equation}
In our case, particles A, B, and C correspond to two bosons (or fermions) and a graviton. Here $E_p$ denotes the energy of one of the fermions or bosons, and $E_\omega$ denotes the energy of the graviton. From~\eqref{eq:int14}, we have the relation,
\begin{equation}
\frac{\rho_\phi}{\phi_0^2}=\frac{m_\phi^2}{k(k-1)}
\end{equation}
With $w_\phi=\frac{k-2}{k+2}$ we can rewrite~\eqref{eq:A5},
\begin{equation}
\frac{d\Gamma_\phi}{dE_\omega}=\frac{(k+2)(k-1)}{2m_\phi^2}\sum_{n=1}^{\infty}n\omega|\mathcal{P}_n|^2\int dE_p |\mathcal{M}'_n|^2
\end{equation}
Substituting the three-body decay amplitude of bosons, we obtain
\begin{equation}
\frac{d\Gamma^{1\to 3}}{d\Eom} = \frac{1}{32\,\pi^3} \left(\frac{\mu}{M_P}\right)^2 (k+2)(k-1)\left(\frac{\omega}{m_{\phi}}\right)^2\left[\sum_{n=1}^{\infty} n^2 |\mathcal{P}_n|^2\left(\frac{(1 - 2 x)^2}{4 x} \right)\right],\quad 0<x<\frac12
\end{equation}
For the fermions case, we obtain
\begin{equation}
\frac{d\Gamma^{1\to 3}}{dE_\omega}=\frac{y^2}{64\,\pi^3} \left(\frac{m_\phi}{M_P}\right)^2(k+2)(k-1)\left(\frac{\omega}{m_{\phi}}\right)^4\left[ \sum_{n=1}^{\infty} n^4 |\mathcal{P}_n|^2 \frac{1 - 2 x}{x} \left(2 x(x - 1) + 1\right) \right]
,\quad 0<x<\frac12
\end{equation}
Here $x=\frac{E_\omega}{n\omega}$. These are differential three-body decay rates by classical field's viewpoint. 

 \section{The dependence of the GW spectrum on the coupling constant}
 \label{Sec:coupling constant}
 \subsection{Bosonic reheating}
To derive the dependence of the frequency at which $\Omega_{\rm GW}$ takes maximum
as well as the amplitude of $\Omega_{\rm GW}$ around that frequency, 
it is first necessary to determine when GWs are produced dominantly.
Given that the peak frequency of gravitons arises due to the energy of inflaton which is $m_\phi$,
the energy density of gravitons having energy about $m_\phi$ produced during one Hubble time at around a time $t$ is estimated as
\begin{equation}
\label{eq:C1}
\Delta \rho_{\mathrm{GW}}\sim \rho_\phi m_\phi \frac{d \Gamma_\phi^{1 \rightarrow 3}}{dE_\omega} 
\bigg|_{E_\omega=m_\phi} H^{-1},
\end{equation}
where all the factors on the right-hand side should be evaluated at the time $t$. 
In the rest of this section, the differential decay rate is always evaluated at $E_\omega=m_\phi$
and we omit the subscript $E_\omega=m_\phi$ for notational simplicity.
Using the time dependence of each quantity as
\begin{equation}
\rho_\phi \propto a^{-\frac{6 k}{k+2}}, \quad 
m_\phi \frac{d \Gamma_\phi^{1 \rightarrow 3}}{dE_\omega} \propto a^{-\frac{3(k-2)}{k+2}}, 
\quad H^{-1} \propto a^{\frac{3 k}{k+2}}
\end{equation}
we obtain
\begin{equation}
\Delta \rho_{\mathrm{GW}} \propto a^{-\frac{6(k-1)}{k+2}}.
\end{equation}
Thus, GWs produced during the Hubble time around $t_1$ will have energy density at later time $t_2$ given by
\begin{equation}
\Delta \rho_{\mathrm{GW}}\left(t_1 \rightarrow t_2\right) \propto a_1^{-\frac{6(k-1)}{k+2}}\left(\frac{a_1}{a_2}\right)^4.
\end{equation}
Here $a_1=a\left(t_1\right)$ and $a_2=a\left(t_2\right)$. Comparing this with the energy density of GWs produced at $t_2$, we have
\begin{equation}
\frac{\Delta \rho_{\mathrm{GW}}\left(t_2\right)}{\Delta \rho_{\mathrm{GW}}\left(t_1 \rightarrow t_2\right)} \propto\left(\frac{a_2}{a_1}\right)^{\frac{-2 k+14}{k+2}}
\end{equation}
Since $a_2>a_1, \Delta \rho_{\mathrm{GW}}\left(t_2\right)>\Delta \rho_{\mathrm{GW}}\left(t_1 \rightarrow t_2\right)$ for $k \leq 6$ and $\Delta \rho_{\mathrm{GW}}\left(t_2\right)<\Delta \rho_{\mathrm{GW}}\left(t_1 \rightarrow t_2\right)$ for $k \geq 8$. Thus, GWs are dominantly produced at around the time of reheating for $k \leq 6$ and at around the beginning of the oscillation state for $k \geq 8$.

Having clarified the time when GWs are dominantly produced depends on whether $k\le 6$ or $k\ge 8$,
let us explore how the peak frequency of GWs at present time depends on $\mu$ for the case $k\leq6$ first.  Assuming there is no entropy production after reheating,
the reheating temperature $T_{\rm rh}$ is related to the scale factor at the time of reheating as
\begin{equation}
\label{eq:C6}
\frac{a_0}{a_\text{rh}}=\frac{T_\text{rh}}{T_0}\left(\frac{g_{\star s}\left(T_\text{rh}\right)}{g_{\star s}\left(T_0\right)}\right)^{1 /3}.
\end{equation}
As it is derived in \cite{Garcia:2020wiy}, the dependence of $T_{\rm rh}$ on $\mu$ is given by
\begin{equation}
T_{\text{rh}}\propto \mu^{\frac{k}{2k-2}}.
\end{equation}
The effective mass of inflaton depends on $\mu$ as
\begin{equation}
m_{\phi, \mathrm{rh}} \propto \mu^{\frac{k-2}{k-1}}.
\end{equation}
Putting the above relations together, the peak frequency at present is evaluated as
\begin{equation}
f_{\mathrm{GW}, 0} \propto m_{\phi, \mathrm{rh}} \frac{a_{\mathrm{rh}}}{a_0} \propto \mu^{\frac{k-4}{2k-2}} \quad (k\leq 6).
\end{equation}
For the case $k \geq 8$, GWs are dominantly produced at around the end of inflation
and the cosmological red-shift must be computed from the inflation end,
\begin{equation}
\label{C:a0-aend}
\frac{a_0}{a_\text{end}}=\left(\frac{a_\text{rh}}{a_\text{end}}\right)\frac{T_\text{rh}}{T_0}\left(\frac{g_{\star s}\left(T_\text{rh}\right)}{g_{\star s}\left(T_0\right)}\right)^{1 / 3}.
\end{equation}
In \cite{Garcia:2020wiy},
the dependence of the reheating time on the coupling constant is derived, and it is given by
\begin{equation}
\frac{a_\text{rh}}{a_\text{end}}\propto \mu^{-\frac{k+2}{3k-3}}.
\end{equation}
We thus obtain
\begin{equation}
\frac{a_0}{a_\text{end}}\propto \mu^{\frac{k-4}{6(k-1)}} 
\end{equation}
The peak frequency at the present time then depends on $\mu$ as
\begin{equation}
f_{\mathrm{GW}, 0} \propto m_{\phi, \text{end}} \frac{a_{\text{end}}}{a_0} \propto \mu^{-\frac{k-4}{6(k-1)}}  \quad (k\geq 8).
\end{equation}

Next, we study the dependence of the height of $\Omega_{\rm GW}$ at the peak frequency
on $\mu$. 
For $k\leq 6$, GWs are primarily produced around the time of reheating, 
so the time of GW production is set to the lifetime of inflaton $1/\Gamma_\phi^{1\to2}$. 
Then, energy density of GWs produced at that time can be estimated as
\begin{equation}
\label{C:boson-k<6}
\rho_\text{GW}(a_\text{rh})\simeq\rho_{\phi}(a_\text{rh})
m_\phi \frac{d \Gamma_\phi^{1 \rightarrow 3}}{dE_\omega} \bigg|_{a=a_{\rm rh}} \times
\frac{1}{\Gamma_\phi^{1\to 2}}
\end{equation}
From this relation, the differential energy density evaluated at $E_\omega=m_\phi$ is given by
\begin{equation}
\left.\frac{d \rho_\text{GW}}{d E_\omega}\right|_{a=a_\text{rh}}\simeq\rho_\phi(a_\text{rh})\frac1{\Gamma_\phi^{1\to 2}}\left.\frac{d\Gamma_\phi^{1\to 3}}{d E_\omega}\right|_{a=a_\text{rh}}.
\end{equation}
The GW spectrum at the present time is related to that in the reheating time as
\begin{equation} \label{eq:oGW}
	\Omega_\text{GW}(f) = \frac{1}{\rho_c}\, \frac{d\rho_\text{GW}}{d\ln f} = \Omega_\gamma^{(0)}\, \frac{d(\rho_\text{GW}/\rho_\text{R})}{d\ln f} = \Omega_\gamma^{(0)}\, \frac{\gs(a_\text{rh})}{\gs(a_0)} \left[\frac{\gss(a_0)}{\gss(a_\text{rh})}\right]^{4/3}\,\frac{d(\rGW(a_\text{rh})/\rR(a_\text{rh}))}{d\ln \Eom(a_\text{rh})}\,.
\end{equation}
From the explicit expressions of the two-body and the three-body decay rates given in the main text,
we find 
\begin{equation}
\frac1{\Gamma_\phi^{1\to 2}}\left.\frac{d\Gamma_\phi^{1\to 3}}{d E_\omega}\right|_{a=a_\text{rh}}E_\omega(a_\text{rh})\propto m_{\phi,\text{rh}}^2.
\end{equation}
Use the definition of reheating $\rho_{\phi, {\rm rh}}=\rho_\text{R,{\rm rh}}$,
we obtain the following scaling law;
\begin{equation}
\Omega_\text{GW}\propto\frac{d(\rGW(a_\text{rh})/\rR(a_\text{rh}))}{d\ln \Eom(a_\text{rh})}\propto m_{\phi,\text{rh}}^2.
\end{equation}
Substituting the dependence of $m_\phi$ on $\mu$, we finally obtain
\begin{equation}
\Omega_\text{GW} \propto \mu^{\frac{2(k-2)}{k-1}}\quad (k\leq 6).
\end{equation}
Here it should be understood that $\Omega_\text{GW}$ means the peak of the GW spectrum.

For $k\geq 8$, given that GWs are primarily produced around the end of inflation,  
we should use $\Delta t=1/H_\text{end}$ as the time interval during which the GWs are primarily generated.
Then, we obtain
\begin{equation}\label{eq:appbg}
\rho_\text{GW}(a_\text{end})\simeq\rho_\phi(a_\text{end})m_\phi \frac{d \Gamma_\phi^{1 \rightarrow 3}}{dE_\omega} \bigg|_{a=a_{\rm end}}\times
\frac{1}{H_\text{end}},
\end{equation}
from which the following relation is obtained
\begin{equation}
\left.\frac{d \rho_\text{GW}}{d E_\omega}\right|_{a=a_\text{end}}\simeq\rho_\phi(a_\text{end})\frac1{H_\text{end}}\left.\frac{d\Gamma^{1\to 3}}{d E_\omega}\right|_{a=a_\text{end}}.
\end{equation}
With the red-shift relation of $\rho_\text{GW}$,
\begin{equation}\label{eq:C21}
\left.\frac{d \rho_\text{GW}}{d E_\omega}\right|_{a=a_\text{rh}}E_\omega(a_\text{rh})=\left.\frac{d \rho_\text{GW}}{d E_\omega}\right|_{a=a_\text{end}}E_\omega(a_\text{end})\left(\frac{a_\text{end}}{a_\text{rh}}\right)^4,
\end{equation}
it can be found that $\rho_R(a_\text{rh})$ depends on $\mu$ as
\begin{equation}\label{eq:C22}
\rho_R(a_\text{rh})=\rho_\phi(a_\text{rh})=\rho_\phi(a_\text{end})\left(\frac{a_\text{end}}{a_\text{rh}}\right)^{\frac{6k}{k+2}}\propto \mu^{\frac{2k}{k-1}}.
\end{equation}
Combining these results together, we finally obtain 
\begin{equation}
\Omega_\text{GW} \propto \mu^{\frac{4k+2}{3k-3}}\quad (k \geq 8).
\end{equation}

\subsection{Fermionic reheating}
As in the case of the bosonic reheating, we start with the estimation
of the time when dominant GWs are produced.
In Eq.~(\ref{eq:C1}), the time dependence of each quantity in fermionic case is given by
\begin{equation}
\rho_\phi \propto a^{-\frac{6 k}{k+2}}, \quad 
m_\phi \frac{d \Gamma_\phi^{1 \rightarrow 3}}{dE_\omega} \propto a^{-\frac{9(k-2)}{k+2}}, \quad H^{-1} \propto a^{\frac{3 k}{k+2}},
\end{equation}
from which we obtain
\begin{equation}
\Delta \rho_{\mathrm{GW}} \propto a^{\frac{6(3-2k)}{k+2}}.
\end{equation}
Following the same logic as in the bosonic case,
we have
\begin{equation}
\frac{\Delta \rho_{\mathrm{GW}}\left(t_2\right)}{\Delta \rho_{\mathrm{GW}}\left(t_1 \rightarrow t_2\right)} \propto\left(\frac{a_2}{a_1}\right)^{\frac{26-8k}{k+2}}.
\end{equation}
Since $a_2>a_1, \Delta \rho_{\mathrm{GW}}\left(t_2\right)>\Delta \rho_{\mathrm{GW}}\left(t_1 \rightarrow t_2\right)$ for $k \leq 2$ and $\Delta \rho_{\mathrm{GW}}\left(t_2\right)<\Delta \rho_{\mathrm{GW}}\left(t_1 \rightarrow t_2\right)$ for $k \geq 4$. Thus, GWs are dominantly produced at around the time of reheating for $k \leq 2$ and at around the beginning of the oscillation state for $k \geq 4$.

Based on the above observation,
the dependence of the peak frequency of GWs on $y$ should be investigated in two cases: $k = 2$ and $k \geq 4$, separately. 
For the first case $k=2$, we again use the redshift relation~\eqref{eq:C6}.
In \cite{Garcia:2020wiy}, it was found that the reheating temperature $T_{\rm rh}$ depends on $y$ as
\begin{equation}
\label{C:fer-Trh-y-k<7}
T_{\text{rh}}\propto y^{\frac{k}{2}} \quad (k<7).
\end{equation}
Thus, $T_{\rm rh} \propto y$ for $k=2$.
The effective mass of inflaton depends on $\mu$ as
\begin{equation}
m_{\phi, \mathrm{rh}} \propto y^{k-2}\quad (k<7).
\end{equation}
Thus, $m_{\phi, \mathrm{rh}} \propto y^0$ for $k=2$.
The peak frequency at the present time can then be evaluated as
\begin{equation}
f_{\mathrm{GW}, 0} \propto m_{\phi, \mathrm{rh}} \frac{a_{\mathrm{rh}}}{a_0} \propto y^{-1} \quad (k=2).
\end{equation}
For cases $k \geq 4$, GWs are dominantly produced at around the end of inflation
and we should use (\ref{C:a0-aend}).
The dependence of the reheating time on $y$ was obtained in \cite{Garcia:2020wiy} and it is given by
\begin{equation}
\frac{a_\text{rh}}{a_\text{end}}\propto 
\begin{cases}
    y^{-\frac{k+2}3} &  (k < 7)\,,\\
    y^{-\frac{k+2}{k-4}} & (k > 7)\,.
\end{cases}
\end{equation}
Using the dependence (\ref{C:fer-Trh-y-k<7}) and $T_{\rm rh}\propto y^\frac{3k}{2k-8}$ for $k>7$ \cite{Garcia:2020wiy},
we obtain
\begin{equation}
\frac{a_0}{a_\text{end}}\propto 
\begin{cases}
    y^{\frac{k-4}6} &  (k < 7)\,,\\
    y^{\frac{1}{2}} & (k > 7)\,.
\end{cases}
\end{equation}
The peak frequency at the present time is then evaluated as
\begin{equation}
f_{\mathrm{GW}, 0} \propto m_{\phi, \text{end}} \frac{a_{\text{end}}}{a_0} \propto \begin{cases}
    y^{\frac{k-4}6} &  (4\leq k < 7)\,,\\
    y^{\frac{1}{2}} & (k > 7)\,.
\end{cases}
\end{equation}

Lastly, we determine the dependence of the height of $\Omega_{\rm GW}$ on $y$ for two cases: $k = 2$ and $k \geq 4$. 
For $k=2$, GWs are primarily produced around the time of reheating. 
Thus, GW density can be estimated by (\ref{C:boson-k<6}) with the decay rate being replaced
with that for the fermionic decay. 
Following the same calculation procedures as in the bosonic case, we obtain
\begin{equation}
\Omega_\text{GW}\propto\frac{d(\rGW(a_\text{rh})/\rR(a_\text{rh}))}{d\ln \Eom(a_\text{rh})}\propto m_{\phi,\text{rh}}^2.
\end{equation}
Substituting the dependence of $m_{\phi, {\rm rh}}$ on $\mu$, we obtain
\begin{equation}
\Omega_\text{GW} \propto y^0\quad (k=2).
\end{equation}
For $k\geq 4$, GWs are primarily produced around end of inflation, 
so the GW density can be estimated by (\ref{eq:appbg}) with the decay rate being replaced
with that for the fermionic decay.
Following the same calculation procedures as in the bosonic case, we finally obtain 
\begin{equation}
\Omega_\text{GW} \propto 
\begin{cases}
    y^{\frac{-2k+14}{3}} &  (4\leq k < 7)\,,\\
    y^{0}=\text{constant} & (k > 7)\,.
\end{cases}
\end{equation}

\newpage
\bibliographystyle{apsrev}  
\bibliography{references}

\end{document}